\documentclass[twocolumn,twocolappendix]{aastex631}

\usepackage{threeparttable}%
\usepackage{mathtools}
\usepackage[T1]{fontenc}
\usepackage{tabularx}

\shorttitle{Radio-loud Quasars are Mergers}
\shortauthors{Breiding et al.}
\graphicspath{{./}{figures/}}

\begin{document}

\title{Powerful Radio-Loud Quasars are Triggered by Galaxy Mergers in the Cosmic Bright Ages}

\correspondingauthor{Peter Breiding}
\email{pbreiding@gmail.com}\\

\author[0000-0003-1317-8847]{Peter Breiding}
\affiliation{The William H. Miller III Department of Physics \& Astronomy, Johns Hopkins University, Baltimore, MD 21218, USA}

\author[0000-0003-1564-3802]{Marco Chiaberge}
\affiliation{The William H. Miller III Department of Physics \& Astronomy, Johns Hopkins University, Baltimore, MD 21218, USA}
\affiliation{Space Telescope Science Institute for the European Space Agency (ESA), ESA Office, 3700 San Martin Drive, Baltimore, MD 21218, USA}

\author[0000-0003-3216-7190]{Erini Lambrides}
\thanks{NPP fellow}
\affiliation{The William H. Miller III Department of Physics \& Astronomy, Johns Hopkins University, Baltimore, MD 21218, USA}
\affiliation{NASA Goddard Space Flight Center, Greenbelt, MD 20771, USA}

\author[0000-0002-7676-9962]{Eileen T. Meyer}
\affiliation{Department of Physics, University of Maryland Baltimore County, Baltimore, MD 21250, USA}

\author[0000-0002-9895-5758]{S. P. Willner}
\affiliation{Center for Astrophysics \textbar\ Harvard \& Smithsonian, 60 Garden St., Cambridge, MA 02138, USA}

\author[0000-0002-6875-1543]{Bryan Hilbert}
\affiliation{Space Telescope Science Institute, 3700 San Martin Drive, Baltimore, MD 21218, USA}

\author[0000-0002-7284-0477]{Martin Haas}
\affiliation{Astronomisches Institut, Ruhr-University, Bochum, Germany}

\author[0000-0003-2884-7214]{George Miley}
\affiliation{Leiden Observatory, Leiden University, P.O.Box 9513, NL-2300 RA, Leiden, The Netherlands}

\author[0000-0002-3099-1664]{Eric S. Perlman}
\affiliation{Florida Institute of Technology, 150 W. University Blvd, Melbourne, FL 32901, USA}

\author[0000-0002-0106-5776]{Peter Barthel}
\affiliation{Kapteyn Institute, University of Groningen, The Netherlands}

\author[0000-0001-6421-054X]{Christopher P. O'Dea}
\affiliation{Department of Physics and Astronomy, University of Manitoba, Winnipeg, MB R3T 2N2, Canada}

\author[0000-0003-3684-4275]{Alessandro Capetti}
\affiliation{Istituto Nazionale di Astrofisica (INAF)—Osservatorio Astrofisico di Torino, via Osservatorio 20, I-10025 Pino Torinese, Italy}

\author[0000-0003-1809-2364]{Belinda Wilkes}
\affiliation{Center for Astrophysics \textbar\ Harvard \& Smithsonian, 60 Garden St., Cambridge, MA 02138, USA}
\affiliation{H. H. Wills Physics Laboratory, University of Bristol, Bristol BS8 1TL, UK}

\author[0000-0002-4735-8224]{Stefi A. Baum}
\affiliation{Department of Physics and Astronomy, University of Manitoba, Winnipeg, MB R3T 2N2, Canada}

\author[0000-0002-5793-7828]{Duccio F. Macchetto}
\affiliation{Space Telescope Science Institute, 3700 San Martin Drive, Baltimore, MD 21218, USA}

\author[0000-0002-5793-7828]{William Sparks}
\affiliation{Space Telescope Science Institute, 3700 San Martin Drive, Baltimore, MD 21218, USA}
\affiliation{SETI Institute, 339 N Bernado Ave, Mountain View, CA 94043, USA}

\author[0000-0002-5445-5401]{Grant Tremblay}
\affiliation{Center for Astrophysics \textbar\ Harvard \& Smithsonian, 60 Garden St., Cambridge, MA 02138, USA}

\author[0000-0002-5222-5717]{Colin Norman}
\affiliation{The William H. Miller III Department of Physics \& Astronomy, Johns Hopkins University, Baltimore, MD 21218, USA}
\affiliation{Space Telescope Science Institute, 3700 San Martin Drive, Baltimore, MD 21218, USA}



\begin{abstract}

While supermassive black holes are ubiquitous features of galactic nuclei, only a small minority are observed during episodes of luminous accretion.  The physical mechanism(s) driving the onset of fueling and ignition in these active galactic nuclei (AGN) are still largely unkown for many galaxies and AGN-selection criteria.  Attention has focused on AGN triggering by means of major galaxy mergers gravitationally funneling gas towards the galactic center, with evidence both for and against this scenario.  However, several recent studies have found that radio-loud AGN overwhelmingly reside in ongoing or recent major galaxy mergers.  In this study, we test the hypothesis that major galaxy mergers are important triggers for radio-loud AGN activity in powerful quasars during cosmic noon ($1\lesssim z \lesssim2$).  To this end, we compare \textit{Hubble Space Telescope} WFC3/IR observations of the $z>1$ 3CR radio-loud broad-lined quasars to three matched radio-quiet quasar control samples.  We find strong evidence for major-merger activity in nearly all radio-loud AGN, in contrast to the much lower merger fraction in the radio-quiet AGN.  These results suggest major galaxy mergers are key ingredients to launching powerful radio jets.     Given many of our radio-loud quasars are blue, our results present a possible challenge to the ``blow-out’’ paradigm of galaxy evolution models in which blue quasars are the quiescent end result following a period of red quasar feedback initiated by a galaxy merger.  Finally, we find a tight correlation between black hole mass and host galaxy luminosity for these different high-redshift AGN samples inconsistent with those observed for local elliptical galaxies.

\end{abstract}



\section{Introduction} \label{sec:intro}

Essentially all massive galaxies seem to harbor a supermassive black hole (SMBH, taken to be black holes with masses $\mathrm{M_{BH}\gtrsim10^{5}M_{\odot}}$) at their dynamical centers   \citep{kormendy_and_ho_13}.  Yet only a small fraction of these SMBHs are typically observed during a phase of luminous accretion characteristic of active galactic nuclei (AGN).  The unified model for AGN stipulates that accretion of matter onto the SMBH is the primary energy source for the observed radiative output from these systems \citep[see e.g. reviews by ][]{antonucci93,netzer15}.  The accretion of gas and dust (either directly, or indirectly through the tidal stripping of stars), and subsequent viscous dissipation within the developing accretion disk, allows for the conversion of gravitational potential energy into thermal optical-UV disk radiation.  The accretion disk can outshine the host galaxy at visible wavelengths and its ionizing continuum can excite atomic tranisitions in gas clouds both near and far from the SMBH, giving rise to the so-called broad and narrow emission lines seen in type I and II AGN, respectively\footnote{The full-width at half maximum (FWHM) of the broad lines is $\mathrm{\gtrapprox~1,000~km~s^{-1}}$,  originating $\sim$ 0.1 pc from the SMBH. }.  

However, uncertainty persists on how AGN are produced in the context of galaxy evolution and dynamics, where there are various physical processes which might contribute to supplying SMBHs with sufficient gas densities for the initiation of AGN activity.  A large focus in the literature is on whether major galaxy mergers might be the dominant AGN triggering mechanism (especially at high AGN luminosities) through galaxy-wide, gravitationally-induced torques which drive gas towards the galactic center (as bolstered by various semi-analytic and numerical works,
 e.g., \citealt{hopkins+06}).  Curiously, there have been a substantial number of studies which both seem to support \citep[e.g.,][]{koss+10,ellison+11,hong+15,weston+17,goulding+18,gao+20} and oppose \citep[e.g.,][]{cisternas+11,kocevski+12,karouzos+14,villforth+19,lambrides+21_llagn} the relevance of major galaxy mergers in triggering AGN.  Collectively, these conflicting results suggest that the physical properties of both AGNs and their host galaxies, and extragalactic environments, may be pertinent when considering the role of galaxy mergers towards their initial triggering (or even sustained fueling).  Additionally, they highlight the need to look at alternatives to major galaxy mergers for transporting gas from kpc to sub-pc scales in galactic nuclei and ultimately triggering AGN, namely: internal ``secular'' processes, minor/satelite-galaxy accumulation \citep[e.g.,][]{hernquist+95}, or gas inflows induced from other interactions within a larger-scale galaxy cluster environment besides galaxy mergers.  
 
 For AGN residing in clusters, ram-pressure induced triggering \citep{marshall+18}, ``galaxy harrassment'' from high-speed galaxy encounters \citep{moore+96}, and tidal interactions between the host galaxy and gravitational potential of the cluster \citep{byrd+90} are all potential AGN-triggering mechanisms apart from major galaxy mergers.  Among secular processes, common channels for SMBH growth include stellar winds (\citealt{davies+07,vollmer+08}, although see \citealt{dubois+15} who showed supernovae can inhibit black hole growth in low-mass systems), chaotic and cold accretion streams  driven by turbulence in the host galaxy intersellar medium (ISM) \citep[][]{hobbs+11,gaspari+13}, and gravitational disk instabilities mediated by stellar bars (and spiral density waves) \citep[e.g.,][]{shlosman+89}, or dense, inhomegeneous clumps formed through cold accretion streams from the larger-scale halo environment \citep[the so-called ``violent disk instabilities'' which seem to be more efficient at higher redshifts, see e.g.,][]{dekel+09,bournaud+11,gabor+13}.

  
Because the triggering of AGN is 
intimately linked to the growth of SMBHs, studies of AGN triggering will help us better understand the tight scaling relations observed between SMBH mass and various host galaxy properties, typically of the bulge or ``spheroid'' component \cite[e.g.,][]{magorrian+98,gebhardt+00,ferrarese+00,marconi+03,haring+04,gultekin+09,graham+11,mcconnell+13}.  The origin of these correlations is still debated, but leading non-mutually exclusive hypotheses include the hierarchical buildup of both black holes and bulges through galaxy mergers \cite[][]{peng07,jahnke+11}, stellar feedback through starbursts \citep[][]{norman+88,davies+07,wild+10}, and AGN feedback \cite[ e.g.,][]{fabian12,heckman+23}.  AGN feedback can take the role of self-limiting SMBH growth through the ejection or heating of gas by radiatively-coupled outflows or jets \cite[e.g.,][]{nesvadba+10,cuoto+23}, or positive feedback where SMBH growth is coupled to host star formation by compressing the ISM \cite[e.g.,][]{silk13}.  Perhaps the most striking example of AGN feedback is the heating of the intracluster medium (ICM) by large-scale radio jets hosted by massive cD galaxies \citep[see e.g. the famous example of the Perseus cluster,][]{fabian+03,fabian+06}

Several previous studies have found strong evidence for enhanced incidence of major galaxy mergers among samples of radio-loud AGN \citep[][]{heckman+86,colina+95,almeida+12,ivison+12,kaviraj+15,chiaberge+15,noirot+18}, and in constrast to the less conclusive results for AGN samples selected by other criteria \citep[e.g.,][]{grogin+05,gabor+09,georgakakis+09,rosario+15,lambrides+21_llagn,sharma+21}.  Understanding why galaxy mergers seem to be ubiquitous among radio-loud AGN, and not among other AGN samples, may help give us better insight into the dyanamical (co)evolution between SMBHs, AGN, and their host galaxies.  Furthermore, studying the link between galaxy mergers and radio-loud AGN may help give clues to the jet formation physics operating in these systems.

The main goal of this study is to test whether major galaxy mergers are an important process for the triggering of radio-loud AGN.  To this end, we analayzed the host galaxy stellar morphologies of the $z>1$ (all but two $1<z<2$) 3CR sample of broad-lined radio-loud quasars with \textit{Hubble Space Telescope (HST)} WFC3/IR images.  This work is a follow-up study to that of \cite{chiaberge+15}, who analyzed the host galaxies of twelve type~II radio galaxies drawn from the same $z>1$ 3CR parent sample, observed with \textit{HST} WFC3/IR SNAP observations (program SNAP13023).  \cite{chiaberge+15} found much higher merger fractions among radio-loud AGN in comparison to matched samples of inactive galaxies and radio-quiet AGN $-$ essentially all $z>1$ 3CR radio galaxies were mergers in comparison to a merger fraction of $\sim40\%$ for the control samples.  In this work, we focus our analysis on the 29\footnote{We note here that the 3CR quasar 3C~119 was dropped from the merger analysis due to a nearby contaminating point source, and is discussed further in Appendix~section\ref{sec:3c119_dropped}.} type~I (broad-lined) quasars 
from the $z>1$ 3CR sample, in order to complement the type~II radio galaxies studied in \cite{chiaberge+15}.  The observations require \textit{HST} point spread function (PSF) subtractions of the unresolved quasar light in order to examine the host galaxy stellar morphologies, as we discuss further in section~\ref{galfit}.  We then used human experts to blindly classify \textit{HST} WFC3/IR images of our 3CR sample in addition to three matched radio-quiet quasar\footnote{Here we use the term quasar interchangeably between radio-loud and radio-quiet sources.} control samples in order to investigate the role of galaxy mergers in triggering the formation of powerful radio jets. 

Throughout this paper we adopt a $\Lambda$CDM cosmology, with H$_{0}=67.74$ km s$^{-1}$ Mpc $^{-1}$, $\Omega_{\lambda}=0.69$, and $\Omega_{m}=0.31$ \citep{planck+16}.  All magnitudes presented are AB magnitudes \citep{oke_and_gunn83}, and extinction-corrected following the reddening maps of \cite{schlafly+11}.

\section{Sample Properties}
\label{control_samples}

The $z>1$ 3CR sample we consider comprise 64 objects observed in the northern hemisphere in Cambridge, United Kingdom \citep[][]{spinrad+85}.  The 3CR sample is a low-frequency, flux-limited sample ($\mathrm{F_{178\ MHz}>9}$~Jy) of extragalactic objects ($\mathrm{|b|>10^{\circ}}$ and declinations above $10^{\circ}$), unbiased in jet orientation because the radio emission at such low frequencies is dominated by the isotropic lobe contribution.  These 3CR galaxies host relatively young radio jets \citep{murgia+99,dallacasa+21,odea+21}, with edge-brightened FR type~II morphologies \citep[][]{fanaroff_filey_74}, and include some of the most powerful radio-loud AGN known.  Furthermore, the $z>1$ 3CR quasars reside in the so-called ``Bright Ages'' ($ \mathrm{1\lesssim z \lesssim 2}$, an epoch also referred to as ``cosmic noon''), thought to correspond to the cosmic peak of AGN activity, star formation, and galactic bulge assembly \citep[e.g.,][]{alberts+16,hopkins+06}, making them an indispensible sample for examining the relationship between SMBHs and their host galaxies.

\begin{table*}
	\centering
	\caption{\textit{HST} WFC3/IR Observations} 
	\label{table:hst_obs}
	\begin{threeparttable}
		\begin{tabular}{lccccc} 
			\hline
			Sample&\textit{HST}&Pivot&Project&Exposure$^{\dagger}$&5$\sigma$~Sensitivity\\
			&Filter&Wavelength&ID&Time&Limit\\
			&&(nm)&&(s)&(mag/arcsec$^{2}$)\\
			\hline
			3CR&F140W&1392.28&GO-16281&$3\times$300&23.9\\
			M16&F160W&1536.91&SNAP-12613&$4\times$400&24.3\\
			V17&F160W&1536.91&GO-13305&$9\times$250&24.4\\
			M19&F160W&1536.91&GO-14262&$6\times$400&24.5\\
			\hline
		\end{tabular}
		\begin{tablenotes}
			\item[$\dagger$] The exposures are given in the format of number of dithers $\times$ single frame exposure time.
		\end{tablenotes}
	\end{threeparttable}
\end{table*}

\begin{figure}
	\includegraphics[width=\linewidth]{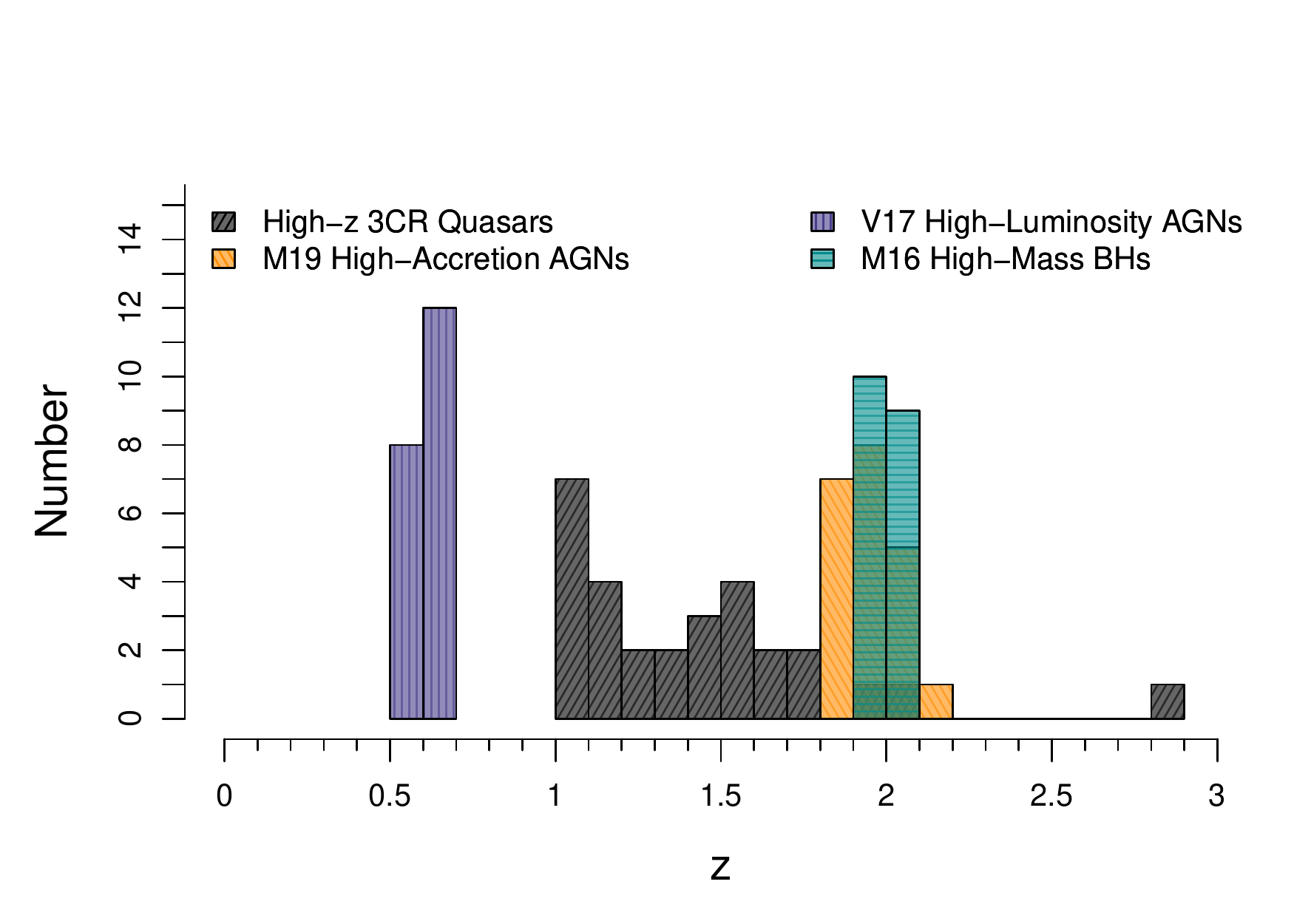}
	\caption{Redshift histogram  distributions for our samples.  The shaded black histogram represents the 3CR sample, orange represents the M19, purple the V17, and cyan the M16 sample.  See section~\ref{control_samples} for a description of each of these control samples.}
	\label{fig:redshifts}
\end{figure}

Our radio-quiet quasar control samples were constructed from three separate archival AGN samples of $\sim$~20 sources observed with similar depth and filter \textit{HST} WFC3/IR observations, as shown in Table~\ref{table:hst_obs}. 
All control quasar samples have published papers assessing the merger fraction of the quasar samples relative to matched samples of inactive galaxies \citep{mechtley+16,villforth+17,marian+19}.   None of these studies found evidence for an enhanced merger fraction among the quasars, thus casting doubt on the scenario of merger-driven triggering for those AGN.  The selection criteria of our three control samples are as follows:

\begin{itemize}
	
\item The first control quasar sample is the high-redshift sample of \cite{mechtley+16} ($1.9 \leq z \leq 2.1$), referred to as ``M16'' for the rest of the paper.  This sample was selected from the Sloan Digital Sky Survey (SDSS)~DR5 quasar catalog \citep{schneider+07}.  This sample of 20 broad-line quasars was selected to have high black-hole masses, spanning the range $\mathrm{9.3 \leq log \left(M_{BH}\right) \leq 9.7}$ (as determined from virial estimates based upon the Mg~II broad line from \citealt{shen+11}), and a uniform color-selection algorithm as described by \cite{richards+02}.  Bolometric luminosities from this sample are from \cite{shen+11}.
	
\item The second sample is that of \cite{villforth+17}, which we will refer to as ``V17''.  This sample is comprised of 20 X-ray selected AGN (\textit{ROSAT} X-ray detections) crossmatched with SDSS DR5 optical quasar broad-line detections \citep[][]{anderson+07}.  The redshift range spans $0.5\leq z \leq 0.7$, and the quasars were also selected to have bolometric luminosities in excess of $\mathrm{10^{45}~erg~s^{-1}}$.  Black hole masses and bolometric luminosities for this sample are from \cite{shen+11}.

\item Our final control sample is composed of the $1.8 \leq z \leq 2.2$ broad-line quasars from  \cite{marian+19}, which we refer to as ``M19''.  These were sub-selected from the SDSS~DR7 quasar catalog \citep[][]{schneider+10,shen+11}.  These 21 AGN were selected to have $\mathrm{8.5\leq log\left( M_{BH}\right)\leq 8.7}$ (as determined by the Mg~II broad-line virial estimates from \citealt{shen+11}), and more importantly Eddington ratios $>0.7$.  Here the Eddington ratio, $\lambda$, is defined as $\mathrm{\lambda \equiv L_{bol}/L_{Edd}}$, where $\mathrm{L_{bol}}$ is the bolometric luminosity, and $\mathrm{L_{Edd}}$ is the Eddington luminosity\footnote{Specifically, the Eddington luminosity is the transition point at which the outward radiation pressure resulting from electron scattering with accretion disk photons overcomes the gravitational force from the black hole, and is usually taken to be $\mathrm{1.26\times10^{38}\left(M_{BH}/M_{\odot}\right)~erg~s^{-1}}$.} describing the limting luminosity needed to halt spherical accretion.  Taken together (and assuming some fixed accretion efficiency), these assumptions imply high mass-accretion rates and thus arguably the most promising selection criteria for constraining the AGN triggering mechanism(s) in these systems.  This sample has the same uniform color-selection criteria as M16.
	
\end{itemize}

Figure~\ref{fig:redshifts} shows the redshift distributions for our four samples.  The 3CR radio-loud AGN fall between the V17 low-redshift ($z\sim0.7$) sample and the two high-redshift ($z\sim2$) samples (M16 and M19).  The control samples have negligible redshift ranges and essentially sample distinct cosmic epochs close to, but generallly on either side, of the distributed 3CR cosmic ages.  The V17 low-$z$ sample has a mean cosmic age of $\sim$~7.3~Gyr, while the high-$z$ M16 and M19 samples have a mean cosmic age of $\sim$~3.3~Gyr.  The mean (and median) cosmic age of our 3CR radio-loud AGN sample is $\sim$~4.6~Gyr, corresponding to $z=1.4$.  While the redshifts of our samples do not exactly match, the control groups effectively sample towards the beginning and right after the cosmic evolution of our 3CR sources.  This gives us some leverage towards discriminating enhanced incidence of galaxy mergers among radio-loud sources while also considering potential redshift evolution in our interpretations.

\begin{figure*}
	\includegraphics[width=\textwidth]{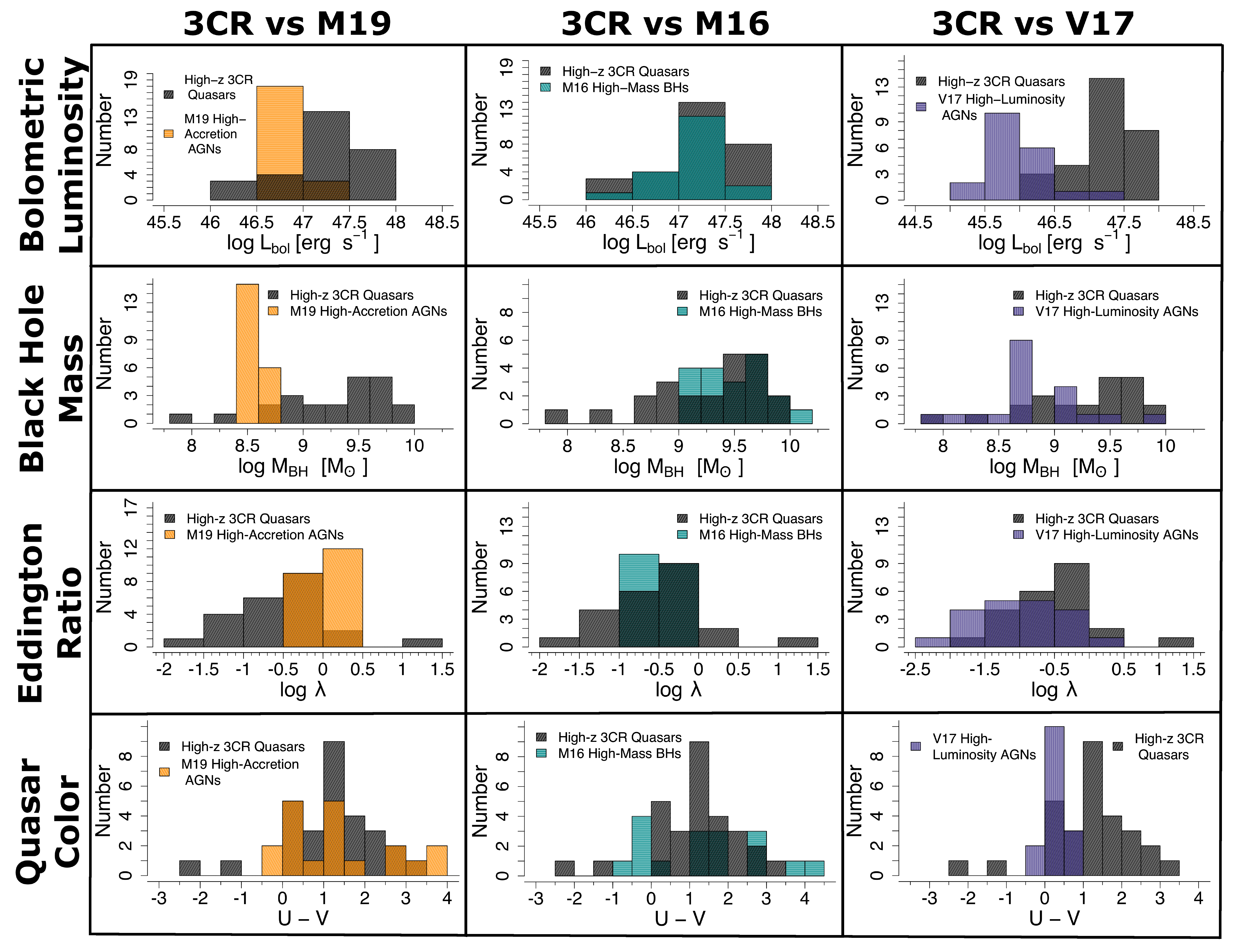}
	\caption{Histogram comparisons of the black hole mass, bolometric luminosity, Eddington ratio, and rest-frame U-V quasar color of the control samples to the radio-loud 3CR sample. The shaded black histograms represent the 3CR sample, orange represent the M19, purple the V17, and cyan the M16 sample.}
	\label{fig:control_hists}
\end{figure*}  

In order to verify that the control samples are all radio-quiet, we cross-matched each against the FIRST, NVSS, and VLASS Very Large Array (VLA) all-sky surveys.  These surveys have the required depth and sky coverage to assess the radio-loudness of our quasars, where we parametrize radio-loudness by the rest-frame ratio of 5~GHz radio to B-band optical flux density after K-corrections\footnote{This was done for the optical data by choosing the appropriate photometric band which redshifts to B-band for our sources (photometry follows the methodology described later in this section when describing quasar colors), and we assumed a flat spectral index (in frequency) for extrapolating the radio.} and use a cutoff of ten to delineate radio-loud from radio-quiet objects \citep{kellerman+94}.  One of the M19 sources, one of the M16 sources, and five of the V17 sources were radio-loud.  We removed these radio-loud sources from our control samples. 

Figure~\ref{fig:control_hists} compares the black hole masses, bolometric luminosities, quasar colors, and Eddington ratios between the 3CR radio-loud AGN and the control radio-quiet samples. For the 3CR, we compiled all black hole massess from different literature sources as given in Table~\ref{table:3cr_properties}.  3CR bolometric luminosities were determined from quasar PSF magnitudes obtained during our \texttt{Galfit} decompositions described further in section~\ref{galfit} and given in Table~\ref{table:3cr_properties}.  Bolometric correction factors were obtained from \cite{azadi+22} for radio-loud quasars at $1<z<2$, after redshifting the F140W filter to the appropriate rest-frame values for each source.  Black hole massess and bolometric luminosities are taken from the reference papers for the control samples; values are given in Table~\ref{table:control_sample_properties}.     Quasar U - V colors were determined in the rest frame, using nominal effective wavelengths of 245\footnote{We use this near-UV wavelength for our ``U band'' as it roughly corresponds to the rest-frame wavelength of our \textit{HST} WFC3/UVIS F606W observations for most of our 3CR sample.}~nm and 580~nm for U and V band, respectively (U - V colors are given in Tables~\ref{table:3cr_properties} and \ref{table:control_sample_properties}).  In order to obtain reliable rest-frame color estimates, we used the appopriate observer-frame photometric band which most closesly redshifts to the rest frame U and V bands.  To this end, we utilize SDSS~DR14 \textit{ugriz} photometry \citep{paris+18}, near-IR \textit{JHK} photometry from the UKIRT Infrared Deep Sky Survey \citep[UKIDSS,][]{lawrence+07}, and the \textit{HST} observations described further in section~\ref{data_analysis}.  Once the appropriate photometric band is chosen, we K-corrected to the nominal rest-frame wavelengths using the near-IR-optical quasar template spectrum from \cite{glikman+06}.  For the SDSS and \textit{HST} data, our quasar colors were constructed using PSF magntiudes in order to avoid any host galaxy contamination (these are not available for UKIDSS data products).     

In general, the distributions shown in Figure~\ref{fig:control_hists} cover similar regions of the overall parameter space.  We also performed two-tailed Kolmogorov–Smirnov (KS) tests to assess whether the control samples and 3CR sample were consistent with being drawn from the same parent population (for all parameters of interest).  The only sample comparisons in which we were able to reject the null hypothesis that the data came from the same population distributions at the 99\% or greater significance level are the M19 vs 3CR black hole masses, M19 vs 3CR Eddington ratios, M16/V17 vs 3CR bolometric luminosities, and V17 vs 3CR quasar colors.  The fact that the KS tests for the M19 vs 3CR black hole masses and Eddington ratios imply different parent distributions can be understood in terms of the M19 selection critera creating artifically narrow distributions.  However, these M19 distributions are still fairly representative of the corresponding 3CR distributions.  Similarly, while the M16 vs 3CR black hole masses and V17 vs 3CR quasar colors have different distributions, the range of values in the control samples is still fairly representative of the 3CR quasars.  However, the V17 bolometric luminosities do seem to generally be lower than the 3CRs by $\sim$~an order of magnitude, even though there is substantial overlap.  Along with the lower redshifts of the V17 sample, this is another factor to consider when comparing the V17 against 3CR merger fractions.  The V17 sample was selected to have high bolometric luminosities (even if they do not approach the levels of the other samples), and were shown in V17 to be in excess of $\mathrm{10^{45}erg~s^{-1}}$ where major mergers are expected to play a dominant role in AGN triggering based on theoretical analytical and simulaiton-based arguments.

\begin{table*}
	\centering
	\caption{3CR Quasar \& Host Galaxy Properties} 
	\label{table:3cr_properties}
	\begin{threeparttable}
		\begin{tabular}{lcccrrcccr} 
			\hline
			Source&z&log~$\mathrm{\left(\frac{M_{BH}}{M_{\odot}}\right)}$&$\mathrm{log~\left(L_{bol}\right)}$&U-V&log~$(\lambda)$&$\mathrm{m_{quasar}}$&$\mathrm{m_{host}}$&$\mathrm{R_{eff}}$&n \\
			Name&&&($\mathrm{erg\ s^{-1}}$)&(mag)&&(mag)&(mag)&(kpc)& \\
			\hline
			3C~2$^{m}$&1.04&8.7$^{c}$&46.5&1.50&$-$0.85&19.3&19.8&10.0&1.7 \\
			3C~9$^{m}$&2.01&9.6$^{b}$&47.8&1.53&$-$0.32&17.5&19.2&4.7&4$^{\dagger}$ \\
			3C~14$^{m}$&1.47&9.4$^{a}$&47.2&2.22&0.04&18.2&19.3&4.8&4$^{\dagger}$ \\
			3C~43$^{m}$&1.46&9.2$^{a}$&46.5&1.06&$-$1.03&20.2&20.6&11.0&2.1 \\
			3C~68.1$^{m}$&1.23&9.9$^{a}$&47.4&2.78&$-0.52$&17.4&19.6&5.3&4$^{\dagger}$ \\
			3C~82$^{n~\dagger\dagger\dagger}$&2.87&...&47.9&1.30&$...$&18.4&...&...&... \\
			3C~119$^{\dagger\dagger}$&1.02&...&47.3&1.09&$...$&17.1&...&...&... \\
			3C~181$^{m}$&1.38&9.6$^{a}$&47.3&0.44&$-0.88$&18.0&19.7&4.3&4$^{\dagger}$ \\
			3C~186$^{m}$&1.07&9.1$^{b}$&47.2&2.20&$-0.56$&17.5&18.7&30.3&1.9 \\
			3C~190$^{m}$&1.20&7.8$^{d}$&47.3&1.23&$1.17$&17.6&19.4&16.6&1.2 \\
			3C~191$^{m}$&1.96&9.0$^{b}$&47.5&1.67&$-0.16$&18.3&19.0&2.8&3.9 \\
			3C~204$^{m}$&1.11&9.7$^{a}$&47.7&0.31&$-0.56$&17.1&19.8&10.3&4$^{\dagger}$ \\
			3C~205$^{m}$&1.53&9.6$^{a}$&48.0&0.57&$-0.02$&16.5&19.7&11.7&4$^{\dagger}$ \\
			3C~208$^{m}$&1.11&10.0$^{e}$&47.3&0.19&$-1.42$&17.5&18.9&2.9&4$^{\dagger}$ \\
			3C~208.1$^{m}$&1.02&...&46.5&0.93&$...$&19.2&19.9&10.9&1.27 \\
			3C~212$^{m}$&1.05&9.3$^{e}$&46.7&2.87&$-1.04$&18.8&19.1&5.9&4$^{\dagger}$ \\
			3C~220.2$^{m}$&1.16&9.4$^{c}$&46.9&1.20&$-1.28$&18.5&19.0&3.8&4$^{\dagger}$ \\
			3C~245$^{m}$&1.03&8.8$^{b}$&47.2&$-$1.33&$-0.41$&17.4&18.9&4.7&4$^{\dagger}$ \\
			3C~268.4$^{m}$&1.40&9.8$^{a}$&47.8&1.11&$-0.32$&16.8&19.0&3.4&4$^{\dagger}$ \\
			3C~270.1$^{m}$&1.53&9.0$^{a}$&47.4&0.47&$-0.17$&18.0&19.4&8.2&4$^{\dagger}$ \\
			3C~280.1$^{m}$&1.66&8.2$^{b}$&47.3&0.02&$0.40$&18.3&20.6&10.8&4$^{\dagger}$ \\
			3C~287$^{n}$&1.05&9.6$^{a}$&47.2&1.60&$-1.02$&17.6&...&...&... \\
			3C~298$^{m}$&1.44&9.6$^{c}$&47.9&0.67&$-0.12$&16.5&17.4&2.8&4$^{\dagger}$ \\
			3C~318$^{m}$&1.57&...&47.1&1.44&$...$&18.6&19.8&5.8&3.3 \\
			3C~325$^{n}$&1.13&...&46.9&1.27&$...$&18.5&18.9&8.7&3.4 \\
			3C~418$^{m}$&1.69&...&47.7&$-$2.19&$...$&17.4&19.0&6.6&4$^{\dagger}$ \\
			3C~432$^{m}$&1.79&9.7$^{b}$&47.6&1.82&$-0.75$&17.8&20.0&5.6&4$^{\dagger}$ \\
			3C~454$^{n}$&1.76&8.9$^{a}$&47.5&3.23&$0.19$&18.0&...&...&... \\
			4C~16.49$^{m}$&1.27&9.8$^{a}$&46.8&2.26&$-1.91$&19.1&19.5&3.5&4$^{\dagger}$ \\
			\hline
		\end{tabular}
		\begin{tablenotes}
			\item Black hole mass measurements are obtained from the following literature sources (corresponding to the superscripts in the $\mathrm{M_{BH}}$ column): (a) \cite{mclure+06} (b) \cite{liu+06} (c) \cite{shen+08} (d) \cite{wu09} (e) \cite{kozlowski17}.
			\item[$\dagger$] These parameters were forced fixed during the 2D brightness distribution fitting.
			\item[$\dagger\dagger$] 3C~119 is dropped from the merger analysis for reasons discussed in Appendix~section\ref{sec:3c119_dropped}.
			\item[$\dagger\dagger\dagger$] 3C~82 is a compact steep-spectrum radio source associated with a faint (V$=$21) object.  The 3C~82 quasar nature was only recently established by \cite{punsley+20} who discovered strongly redshifted broad emission lines at $z=2.87$, rendering it the most distant quasar in the 3CR sample. 
			\item[m] Classified as a galaxy merger by voting consensus.
			\item[n] Classified as a non-merger by voting consensus.
			\item Empty values for $\mathrm{m_{host}}$ indicate host galaxies deemed unresolved by our \texttt{Galfit} fits, in which case we only model the quasar PSF.
		\end{tablenotes}
	\end{threeparttable}
\end{table*}

\begin{table*}
	\centering
	\begin{threeparttable}
		\caption{Control Sample Quasar \& Host Galaxy Properties} 
		\label{table:control_sample_properties}
		\begin{tabular}{llccccrcccr} 
			\hline
			Source&Sample&z&log~$\mathrm{\left(\frac{M_{BH}}{M_{\odot}}\right)}$&$\mathrm{log~\left(L_{bol}\right)}$&U-V&log~$(\lambda)$&$\mathrm{m_{quasar}}$&$\mathrm{m_{host}}$&$\mathrm{R_{eff}}$&n \\
			Name&&&&($\mathrm{erg\ s^{-1}}$)&(mag)&&(mag)&(mag)&(kpc)& \\
			\hline
			J083253\tnote{m}&M19&1.8&8.5&47.2&3.57&$-0.09$&19.0&21.2&8.4&0.75 \\
			J084632\tnote{n}&M19&2.0&8.7&46.8&1.45&0.35&18.0&...&...&... \\
			J091555\tnote{n}&M19&2.1&8.7&46.9&2.66&$-0.02$&18.4&...&...&... \\
			\hline
		\end{tabular}
		\begin{tablenotes}
			\item This table is published in its entirety online in machine-readable format, where we show a sample portion here for guidance regarding its form and content. 
			\item[$\dagger$] These parameters were forced fixed during the 2D brightness distribution fitting.
			\item[m] Classified as a galaxy merger by voting consensus.
			\item[n] Classified as a non-merger by voting consensus.
			\item Empty values for $\mathrm{m_{host}}$ indicate host galaxies deemed unresolved by our \texttt{Galfit} fits, in which case we only model the quasar PSF.
		\end{tablenotes}
	\end{threeparttable}
\end{table*}         

\section{\textit{HST} Observations \& Data Analysis}
\label{data_analysis}

All of our quasar samples were observed with \textit{HST} WFC3 near-IR observations.  Table~\ref{table:hst_obs} gives the project code, exposure time, \textit{HST} WFC3/IR filter, pivot wavelength, and surface brightness sensitivity limits for each sample.  The sensitivity limits were estimated with the WFC3/IR online Exposure Time Calculator (ETC).  For the ETC exposure calculations we used a $2\times2$ pixel extraction area, elliptical galaxy spectrum, median redshift for the sample, and neglected Milky Way extinction.  The \textit{HST} observations of the 3CR sample also included WFC3/UVIS F606W (pivot wavelength of 588.92~nm) observations in the same orbit as the IR channel exposures.  We used UVIS aperture photometry when appropriate for determining quasar colors.  Generally, there is fairly low host contamination in the UVIS channel because it probes rest-frame near-UV for the 3CR sample, thus allowing for robust quasar magnitude measurements from aperture photometry without PSF-fitting (except for a few cases of star-forming regions, which we masked when measuring fluxes).  

All samples were observed with $\geq$~3 point dithers so as to remove hot pixels/cosmic rays and allow for improved resolution in the final sub-pixel imaging.  All observations were non-destructively read out in a log-linear STEP sequence, intended to allow for high-dynamic range imaging.  The  exposure times for all samples allow for similar surface brightness detection thresholds, although the control samples are \textit{slightly} deeper than the 3CR.  We are testing the hypothesis that the 3CR radio-loud AGN have a higher incidence of galaxy mergers, and greater sensitivity to low surface-brightness merger indicators in the control groups only serves to make our results more robust in the event of a higher 3CR merger fraction.

The 3CR sample was observed with the F140W wide filter (JH gap), with pivot wavelngth of 1392.3~nm.  At the median redshift of the 3CR sample ($z=1.4$), this corresponds to a rest-frame wavelength of $\sim$~580~nm or V-band.  The control-group samples were all observed with the wide H-band F160W filter, corresponding to a pivot wavelength of 1536.9~nm.  For the V17 sample, this corresponds to a rest-frame wavelength of $\sim$~904~nm or Z-band in the near-IR.  For the high-redshift M16/M19 samples, this correpsonds to a rest-frame wavelength of $\sim$~512~nm or V-band.  The V17 observations were chosen to sample $\sim1~\mu$m rest-frame wavelengths in order to serve as a good proxy for stellar mass.  The host galaxy morphologies should not appear substantially different between this wavelength and visible wavelengths, nor should the hosts be substantially contaminated by emission lines at these wavelengths, even if they are especially star-forming \citep{martins+13}.  As shown by \cite{hilbert+16}, the host galaxies of our 3CR sample should also not be substantially contaminated by emission lines, and the same is true for the $z\sim2$ M16/M19 samples which overlap in redshift with the 3CR.

\subsection{Data Reduction}

All \texttt{flt} image files were downloaded from the Mikulski Archive for Space Telescopes (MAST) after having first been processed through the \texttt{calwf3} data reduction pipeline \citep{dressel22}.  We combined the dithered \texttt{flt} image files together with  \texttt{Astrodrizzle} \citep{gonzaga+12}, which also subtracts the sky background and makes corrections for cosmic ray artifacts and geometric distortion.  \texttt{Astrodrizzle} relies on the ``drizzle'' algorithm, or variable-pixel linear reconstruction, in which pixels in dithered CCD images are mapped onto a sub-sampled output image after taking into account rotational and translational shifts between dithered input images and camera distortion \citep{fruchter+02}.  We used a \texttt{final\_pixfrac} parameter of 0.8 (describing the drop size of input pixels) and a \texttt{final\_scale} of 0.06\arcsec for the final pixel scale of the output image.  After experimentation, we found this combination of \texttt{final\_pixfrac} and \texttt{final\_scale} allowed for the best compromise between final image resolution, correlated noise properties, and sensitivity to low-surface-brightness features.  This 0.06\arcsec per pixel final plate scale of our output images oversamples the point spread function (PSF) by a factor of two for the F140W and F160W filters of \textit{HST's} WFC3 camera.      

\subsection{Galfit Decompositions}
\label{galfit}

\begin{figure*}
	\includegraphics[width=\textwidth]{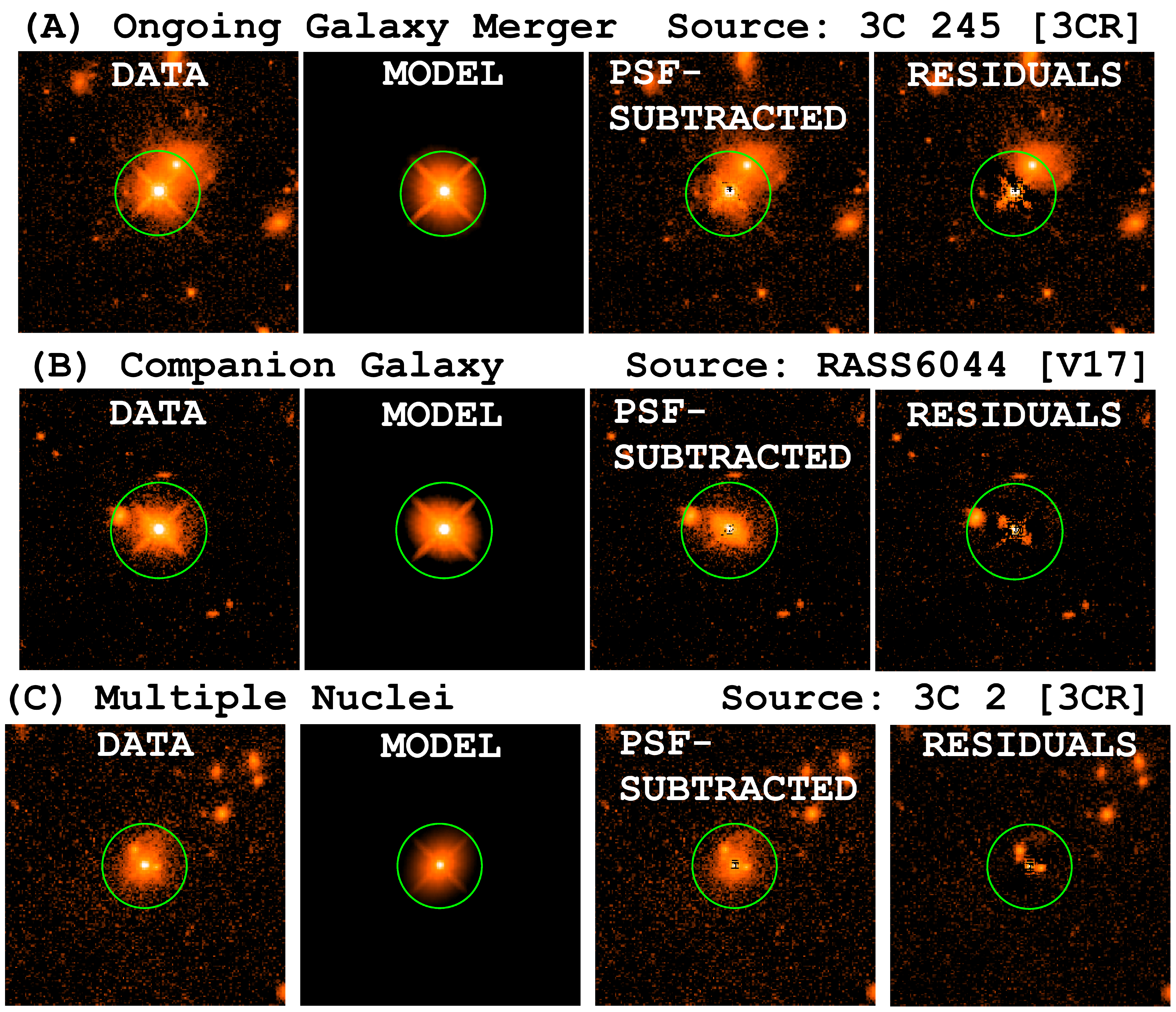}
	\caption{\texttt{Galfit} decomposition panels used for classifications, where we show examples corresponding to each classification category as supported by the expert voting results.  Images are constructed in the original detector frame, with varying right ascension/declination axis orientations.  The classification is given in the upper left of each panel (as supported by the voting results), and source/sample are given in the upper right.  All panel decompositions can be found as online-only figures.  The green circles have radii corresponding to 25~kpc projected onto the plane of the sky for each quasar redshift.}
	\label{fig:panel1}
\end{figure*}  

\begin{figure*}
	\includegraphics[width=\textwidth]{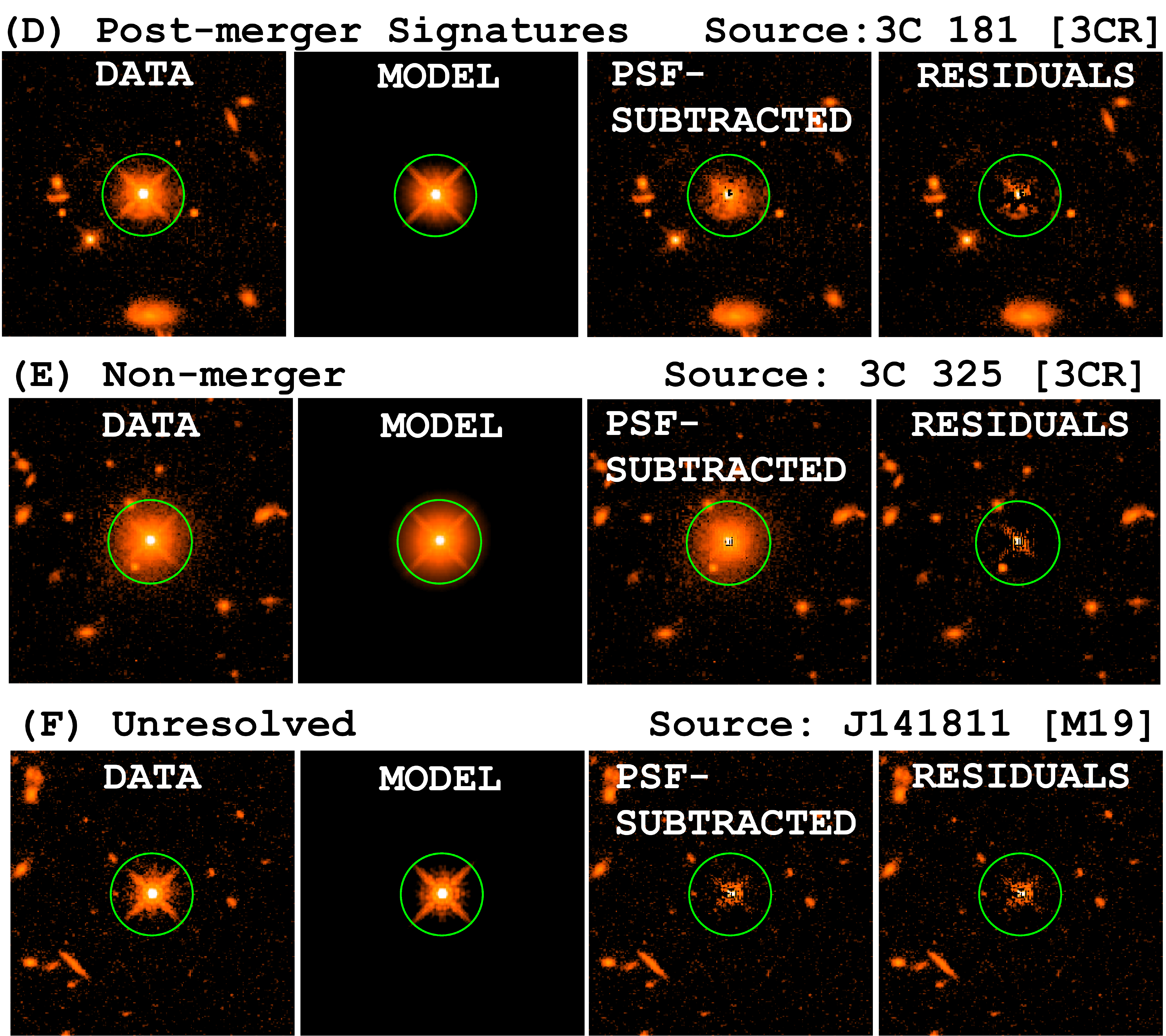}
	\caption{Continuation of Figure~\ref{fig:panel1}.}
	\label{fig:panel2}
\end{figure*} 

After drizzling we performed 2D modeling of the brightness distribution within our images using the \texttt{Galfit} \citep[][]{peng+10} software package.  For simplicity and consistency, we used single S\'ersic profiles to model the host galaxy light distrbution \citep{sersic63,sersic68}, which have the following radial dependence:

\begin{equation}
\mathrm{	I(R)=I_{e}~exp\left[-\kappa \left(\left(\frac{R}{R_{e}}\right)^{1/n}-1\right)\right] }.
\end{equation}

Here, I is the intensity as a function of radius (R), $\mathrm{R_{e}}$ is the effective radius containing half the flux, $\mathrm{I_{e}}$ is the intensity at $\mathrm{R_{e}}$, n is the S\'ersic index, and $\kappa$ is a parameter depending only on n.  Thus, n solely determines how centrally concentrated the light profile is.  Profiles with $\mathrm{n=1}$ correspond to exponential functions common to modeling galaxy disks \citep{freeman70}, and $\mathrm{n=4}$ describes the de Vaucouleurs profile used to model classical bulges \citep{devaucouleurs48}.  

While the S\'ersic profile may not perfectly describe the light distributions of our galaxies, to the first order it is efficient in modeling the underlying brightness distribution of the non-disturbed host galaxy components without the worry of over-fitting.  Since \texttt{Galfit} optimizes model parameters simultaneously when minimizing $\chi ^{2}$ through the non-linear ``Levenberg–Marquardt'' algorithm, this simple host light model helps keep us from oversubtracting the quasar PSF contribution described further below.  

In addition to S\'ersic profiles for the host galaxies, we modeled the sky background, and AGN point sources using synthetic PSFs constructed from the \texttt{Tiny Tim} software package \citep{krist+11}.  Our PSF model was constructed using the near-IR-optical quasar template spectrum from \cite{glikman+06}, redshifted appropriately for each source.  We built PSF models separately for each dithered \texttt{flt} frame, using the WFC3 focus model from the following URL maintained by STScI: \texttt{focustool.stsci.edu/cgi-bin/control.py}. Each PSF was constructed at the chip pixel location of peak flux measured in that \texttt{flt} frame, using a 7.5$^{\prime\prime}$ diameter.  After constructing each PSF model for the different dithered frames, we combined the PSFs by inserting them into blank \texttt{flt} frames and drizzling using the same parameters as our science images, except with a $2\times$ finer plate scale of 0.03$^{\prime\prime}$ per pixel.  This yielded PSF models sampled to a $\sim4\times$ finer pixel scale than the PSF FWHM.  We evaluated the quality of our PSF-subtraction technique for one of our quasars with an unresolved host galaxy in Appendix section~\ref{psfsub_uncertainties}.  We show all of the \textit{HST} WFC3/IR images of our 3CR quasars before and after PSF-subtraction in Appendix section~\ref{sec:3CR_images}.  Similar use of drizzled \texttt{Tiny Tim} PSFs to subtract quasar light for host galaxy morphological analysis in other studies showed good results \citep[][]{villforth+17,zakamska+19}.  

We attempt fitting each source with both a S\'ersic and PSF model component, allowing all relevant parameters to be free unless constraining some to physically motivated values helped improve the fit as determined by $\chi^{2}_{\nu}$ (or allowed the fit to converge to physically meaningful values at all).  The model parameters for the S\'ersic profiles are the index n, effective radius $\mathrm{R_{e}}$, the axis ratio (semi-minor/semi-major axis), position angle, sky position, and integrated magnitude.  For the PSF model, only the sky position and magnitude were free parameters, and the sky background was fit as a constant.  We also used a hand-constructed bad-pixel mask for nearby sources and irregular host galaxy features so as to most accurately model the non-disturbed host and quasar components.  Appendix section~\ref{sec:blended_components} compares this method of masking all pixels containing significant deviations from the quasar and its undisturbed quasar host galaxy component with an alternative method where we include a greater number of model components for very nearby or blended companion galaxies.  The difference in magnitude between the two methods is less than the uncertainties found by \cite{simmons_urry_08} for quasar host galaxies in simulations of \textit{HST} ACS data for the GOODS survey (when comparing quasars with similar fluxes and flux ratios between the quasar and its host galaxy).  For some cases of very bright quasars, we also masked the central few brightest pixels in order to account for non-linear pixel response where the PSF model would not match the saturated pixels.  Tables~\ref{table:3cr_properties} and \ref{table:control_sample_properties} report the best-fit results of our \texttt{Galfit} decompositions.  

In cases where the host galaxy appears to be marginally resolved or unresolved, we used a fixed S\'ersic index of four and placed the S\'ersic component at the same location as the PSF after optimizing a PSF-only fit.  If the S\'ersic effective radius was less than 3 pixels, equivalent to 0.18\arcsec or $\sim$50\% beyond the PSF FWHM, then we considered the host unresolved and used only a PSF-fit in our \texttt{Galfit} galaxy model.  These cases correspond to null values for $\mathrm{m_{host}}$ in Tables~\ref{table:3cr_properties} and \ref{table:control_sample_properties}. 



\section{Classification Methodology}
\label{classifications}

For consistent comparison of our results to those of \cite{chiaberge+15}, who analyzed the host galaxies of the $z>1$ type~II 3CR radio galaxies, we used six human experts for our morphological merger classifications corresponding to six of the authors on this paper.  This has the benefit of taking a more holistic view of galaxy mergers in comparison to quantitive, non-parametric measures of disturbance such as the Gini coefficient, second-order moment of the
brightest 20\% of the light ($\mathrm{M_{20}}$), or asymmetry measure \citep{schade+95,lotz+04}.  These quantitative estimates of disturbance can be effective at measuring post-galaxy-coalescence signatures (i.e., multiple nuclei, tidal features, and other  large-scale asymmetries, see e.g., \citealt{lotz+11}), but would miss many ongoing/incipient galaxy mergers or those with faint signatures that the human eye is more sensitive to.  All measures are likely to be biased by the PSF-subtraction uncertainties, which are inherent to our classification images, as pointed out by previous WFC3/IR quasar merger studies \citep[i.e.,][]{glikman+15,zakamska+19}.

Each human classifier examined the entire image set before final source classifications were decided upon to ensure internal self-consistency.  Each expert classified each quasar image presented in the form of \texttt{Galfit} panel decompositions. Figures~\ref{fig:panel1} and \ref{fig:panel2} show examples.  These panels were constructed based upon our \texttt{Galfit} 2D modeling described in section~\ref{galfit}, and show the original image data, our best-fit \texttt{Galfit} models, the PSF-subtracted images, and the residuals.  As described in section~\ref{galfit}, for unresolved objects for which we could not obtain a physically meaningful S\'ersic fit, we only used a PSF model, and the residuals are the same as the PSF-subtracted image.  Otherwise, the residuals are the data minus the best-fit PSF and S\'ersic models.  Including the residuals in the classifications helps to highlight the faint merger signatures (i.e., global asymmetries and tidal disturbances indicative of a recent or ongoing major galaxy merger), because the host galaxy components consistent with symmetric S\'ersic profiles are removed.  The images were stretched to highlight faint morphological features necessary for careful inspection and classification.  
Each panel was assigned a random number, and the experts had no identifying knowledge for each panel, so these are completely blind classifications.  We classified images according to the following criteria:

\begin{itemize}
	\item \textbf{A: Ongoing galaxy merger} $-$ these show two (or more) massive galaxies with clear signs of interaction such as tidal bridges, streams, shells, or tails.  Cases with unresolved host galaxies that fit this category are still classified as ongoing mergers.   
	\item \textbf{B:  Close companion(s)/incipient galaxy merger} $-$  these show a massive galaxy companion within the 25~kpc radius surounding the quasar but otherwise do not show any signs of interaction.  Major/minor incipient galaxy mergers are further distinguished after classification by a flux ratio of 1:4 (i.e., minor incipient mergers are those where the less massive galaxy is $\geq4\times$ less massive).  Minor incipient mergers were reclassified as non-mergers.     
	\item \textbf{C: Multiple nuclei} $-$ these show a common envelope around two or more nuclei, either stellar or unresolved quasar-like nucei.  
	\item \textbf{D: Post-merger Signatures} $-$  these do not have galaxy companions within 25 kpc but have host galaxies which show light profile asymmetries, tidal features, or other disturbances indicative of a post-coalescence major galaxy merger.  
	\item \textbf{E: Non-merger} $-$ these appear as smooth and symmetric host galaxies without any massive companions within 25 kpc. 
	\item \textbf{F: Host galaxy unresolved/non-detected or PSF-dominated residuals} $-$ for these, the classifier does not identify enough host galaxy component to classify it as any of the above letters.  It may be some combination of very bright PSF, general PSF mismatch, compact host galaxy, undermassive host galaxy, high-redshift cosmological surface brightness dimming, or other factors that lead to this classification. 
\end{itemize}

Figures~\ref{fig:panel1} and \ref{fig:panel2} show examples firmly classified into each of the above categories.  Each classifier's votes were consolidated into the categories of ``merger'' or ``not merger'', where categories of A$-$D counted towards ``merger'' and E/F were counted towards ``not merger''.  This choice to reclassify unresolved sources as non-mergers implicitly assumes we would have seen evidence of a merger if it were there, regardless of incomplete host galaxy information.  We discuss any potential bias resulting from this choice in Appendix section~\ref{sec:unresolved_host_bias}. In the case of sources classified as F, or ``unresolved'', with detected companions within 25~kpc meeting the 1:4 flux ratio cutoff we use to discrimnate massive companions indicative of an incipient major galaxy merger, we reclassified them as ``B'' after voting and these would be considered mergers (see Appendix section~\ref{sec:flux_ratios} where we explain how this is done for cases without host galaxy fluxes determined from our \texttt{Galfit} decompositions).  Finally, we further consolidated the votes into a consensus classification based upon the majority vote, where ties were broken in favor of ``merger''. 

\section{Results}
\label{results} 

\subsection{Merger Incidence}
\label{bayesian_methodology}

\begin{table}
	\centering
	\caption{Consensus Results \& Merger Fractions} 
	\label{table:merger_results}
	\begin{threeparttable}
	\begin{tabular}{lccr} 
		\hline
		Sample&Merger&Not&$\mathrm{f_{m}}^{\dagger}$\\
		&&Merger&\\
		\hline
		3CR&25&3&$0.87^{+0.09}_{-0.14}$\\
		M16&5&13&$0.30^{+0.21}_{-0.17}$\\
		V17&5&10&$0.35^{+0.23}_{-0.20}$\\
		M19&9&11&$0.45^{+0.21}_{-0.20}$\\
		\hline
	\end{tabular}
	\begin{tablenotes}
		\item[$\dagger$] The error bars represent 95\% credible intervals around the mean values.  These are Bayesian merger fractions determined from the posterior distrbutions for each sample's consensus classification.  
	\end{tablenotes}
	\end{threeparttable}
\end{table}

\begin{figure}
	\includegraphics[width=\linewidth]{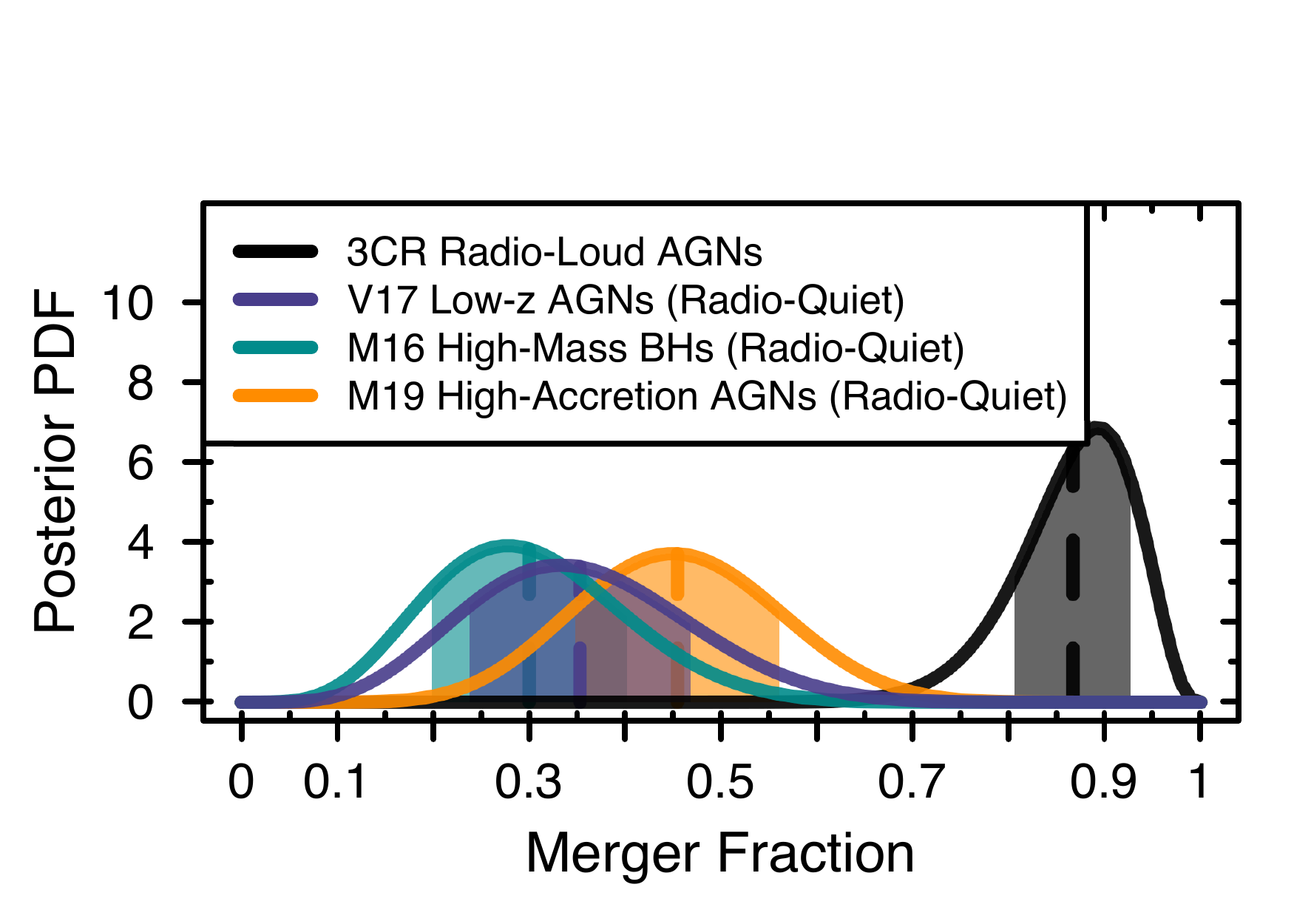}
	\caption{Posterior probability density function (PDF) of the merger fractions for each sample, based upon the consensus classification results given in Table~\ref{table:merger_results}.  Vertical dashed lines with semi-transparent bands show the means of these distributions with 68\% credible intervals.}
	\label{fig:posteriors}
\end{figure}

Table~\ref{table:merger_results} gives the results of our consensus classifications after votes were consolidated into the categories ``merger'' and ``not merger''.  
In order to robustly quantify the 3CR merger fraction and compare to those of the control samples, we used a Bayesian framework as described below.  The consolidated classification of each galaxy as ``merger'' or ``not merger'' turns each observed galaxy into a Bernoulli trial (i.e., ``merger'' would count as a ``success'', and ``not merger'' as ``failure'').  
Therefore, the likelihood of our data is given by the discrete binomial probability distribution:

\begin{equation}
	\label{binomial}
{n \choose k}f_{m}^{k}(1-f_{m})^{n-k}~,
\end{equation}    

where \begin{equation}
	{n \choose k} \equiv \frac{n!}{k!(n-k)!}~.
\end{equation}   
 
Here, $f_{m}$ is the intrinsic merger fraction for a given sample (or probability of any successive Bernoulli trial being a merger), k is the number of mergers observed for that sample, and n is the number of Bernoulli trials observed from that sample.  

We adopted an uninformative uniform proior on merger fraction for each sample, $f_{m}\sim~U[0,1]$.  Combining this prior with the binomial likelihood function given by equation~\ref{binomial},
Bayes' theorem  yields a posterior distribution for merger fraction of $f_{m}\sim~Beta(k+1,n-k+1)$.  For these ``shape parameters'', the Beta distribution is given by:

\begin{equation}
	\begin{split}	
	 Beta(f_{m};k+1,n-k+1) =\\ \frac{\Gamma(2+n)}{\Gamma(k+1)\Gamma(n-k+1)} f_{m}^{k}(1-f_{m})^{n-k}~,
	\end{split}
\end{equation}

 where
 
 \begin{equation}
 	\Gamma(x)\equiv(x-1)!~. 
 \end{equation}

The resulting posterior distributions for  the merger fractions of our different samples based upon the consensus classification results are shown in Figure~\ref{fig:posteriors}.  
A Bayesian two-sample proportion test based upon Monte-Carlo sampling the respective posterior distributions \citep[as implemented in the R package \texttt{BayesianFirstAid},][]{baath2014bayesian} shows that the 3CR sample has a greater intrinsic merger fraction than all control groups at the $>99.9\%$ confidence level.  

\subsection{Monte Carlo Analysis Without Voter Consolidation}
\label{monte_carlo}

\begin{figure}
	\includegraphics[scale=0.4]{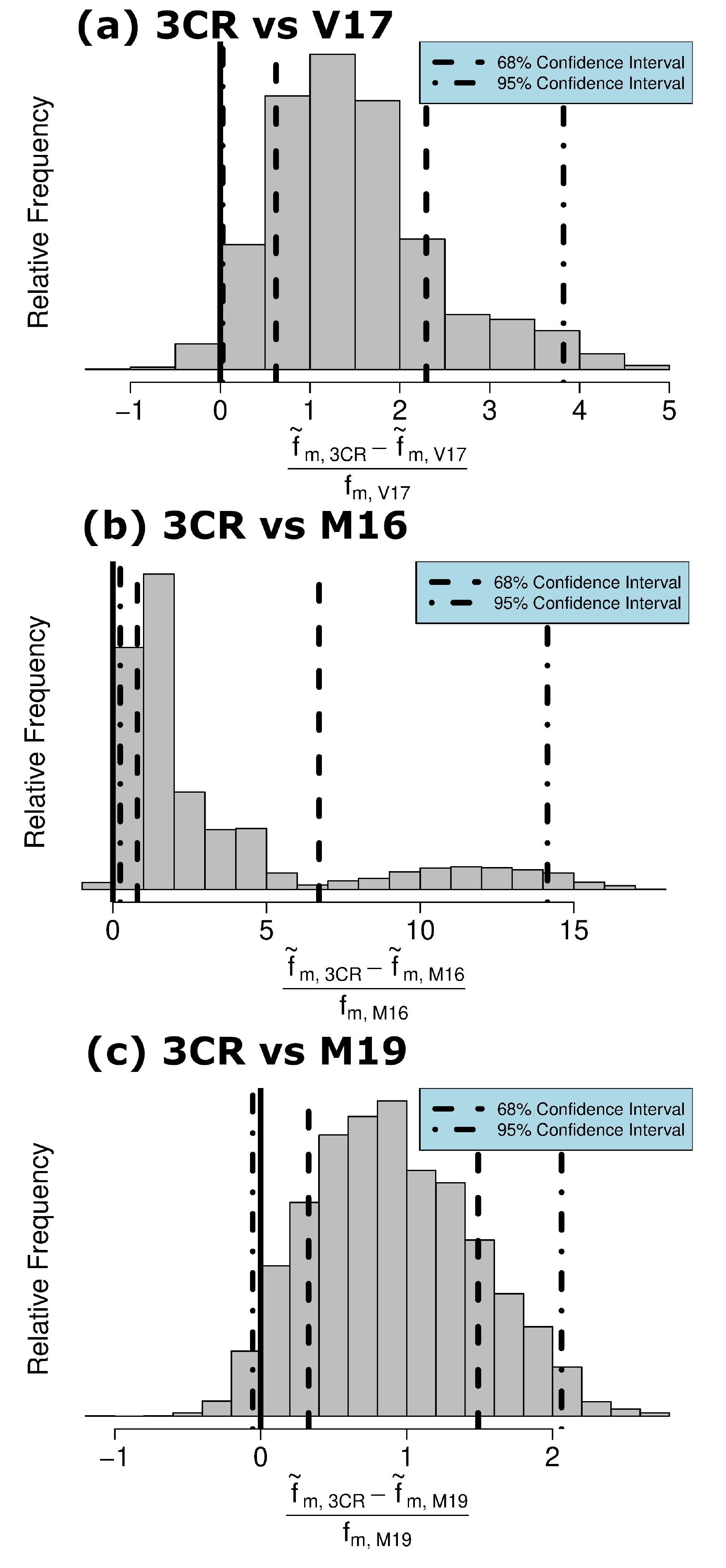}
	\caption{Co-added histograms of each classifer's Monte-Carlo sampled difference in merger fraction of the 3CR sample relative to each control group, scaled by the control sample merger fraction (again defined for each classifier).  The 68\% confidence interval is shown as dashed line, and the 95\% confidence interval as the dot-dashed line.  Positive values indicate an enhnced 3CR merger fraction relative to the given control group.}
	\label{fig:monte_carlo}
\end{figure}  

In order to strengthen the robustness for our result of an enhanced 3CR merger fraction in comparison to the control groups, we performed an additional analysis following a similar methdology to that of \cite{cisternas+11} (see their Section~4.2.1), except within our existing Bayesian framework and using the same six human expert classifications described in section~\ref{classifications}.  In this case, we considered each of the classifier's ``votes'', or classification choices, individually, allowing each classifier to define their own posterior distribution of merger fraction for each sample.  Then, for each control group, we performed a Monte Carlo sampling of the 3CR and control group merger fractions (denoted as $\mathrm{\tilde{f}_{m, 3CR}}$ and $\mathrm{\tilde{f}_{m, CS}}$, respectively) from their posterior distributions.  This was done using 1,000 Monte Carlo samples of $\mathrm{\tilde{f}_{m, 3CR}}$ and $\mathrm{\tilde{f}_{m, CS}}$ per classifier.  We then computed the difference in merger fraction for each classifier from these Monte Carlo samples, scaled by each classifer's control sample merger fraction\footnote{For this, we used the mean of their posterior distribution.}: $\mathrm{\frac{\tilde{f}_{m, 3CR}-\tilde{f}_{m, CS}}{f_{m, CS}}}$.  This allows us to include input from all classifiers on how they perceive the 3CR sample relative to the control groups.  Normalizing the difference in merger fraction, $\mathrm{\tilde{f}_{m, 3CR}}-\mathrm{\tilde{f}_{m, CS}}$,  by each classifier's control sample merger fractions allows for the combination of different classifiers results while adjusting for the difference in personal scale for each classifier.  

Figure~\ref{fig:monte_carlo} shows the results of this analysis for all classifiers after co-adding their histograms.  Positive values correspond to an incidence of higher 3CR merger fraction relative to the given control group being compared. 
Evidence for a statistically significant enhanced merger fraction of the 3CR sample relative to the control groups from this figure would correspond to cases where the confidence interval bounds fall above zero.  As is evident, the 95\% confidence interval bounds for the 3CR vs V17 and 3CR vs M16 sample comparisons falls above zero, as does the 68\% bounds for the 3CR vs M19 sample.  The heavy-tailed nature of  the 3CR vs M16 histogram reflects a greater variance of control sample merger fractions among the classifers. 

This analysis supports the finding of an enhanced merger fraction of the 3CR sample relative to all of the control groups.  However, the statistical significance of our results do appear stronger using the consensus classification methodology outlined in the previous section, rather than this method of combining the votes from all classifiers.  This is consistent with our consensus classification removing some of the human bias inherent to our classifications by our consolidation schema.  If this hypothesis is correct, the inclusion of more classifiers should only serve to bring the consensus classifications closer to the ``truth'' as is implicitly assumed in large-scale voter-based morphological classifications in astronomy such as Galaxy Zoo \citep[][]{lintott+08}. 

\section{Discussion}
\label{discussion}

Our finding of high 3CR merger fraction and evidence for enhanced merger fraction of the 3CR sample relative to the radio-quiet AGN control groups, implies that major galaxy mergers are important to the triggering of radio-loud AGN and the launching of powerful radio jets 
in the cosmic bright ages ($1\lesssim z\lesssim2$).  This result is consistent with that of \cite{chiaberge+15}, who found $\mathrm{f_{m}=0.94^{+0.06}_{-0.16}~}$ for the type~II $z>1$ 3CR radio galaxies.  The finding of high merger fraction in both the type~I and type~II $z>1$ 3CR radio-loud AGN is consistent with AGN unification models which predict AGN type based upon relative orientation with respect to the observer's line of sight \citep{lawrence_elvis_82,antonucci_miller_85,barthel89,antonucci93}.  Given that not all major galaxy mergers host radio-loud AGN, galaxy mergers are consistent with being an important, although not sufficient, process for the launching of radio jets.  

Our result of high merger fraction for the $z>1$ 3CR radio-loud quasars is also consistent with the work of \cite{podigachoski+15}, who find high star-formation rates for the same parent sample of $z>1$ 3CR radio-loud AGN based upon \textit{Herschel} far-infrared and \textit{Spitzer} mid-infrared photometry with corresponding spectral energy distribution decomposition into AGN-heated and star-formation-heated dust components.  \cite{podigachoski+15} interpret these high star-formation rates as likely originating from major gas-rich galaxy mergers, although jet-induced star formation is another possibility and is hard to separate from merger-induced star-formation due to the lack of required spatial resolution in the \textit{Herschel} and \textit{Spitzer} observations.  Follow-up \textit{JWST} mid-infrared observations would help to resolve the dust-enshrouded star-forming regions in the $z>1$ 3CR sample and help better discern the origin of the prodigious, obscured star formation taking place in these systems.

\subsection{Potential Impact of Quasar Environment and Jet-Host Interactions}
\label{confounding_variables}

In selecting our control groups, we focused exclusively on redshift and basic quasar properties.  
The purpose of this selection criteria was to make the presence of a powerful radio jet the major distinguishing variable between the radio-loud and radio-quiet AGN.  However, we made no attempt to further control for quasar environment, as is the case in most other quasar merger studies.  In principle, the large-scale density of group or cluster members, dark matter halo properties, intergalactic medium (IGM), and various host galaxy properties (e.g., stellar mass, gas/dust density and temparature, color, metallicity, size, etc.) are possible confounding variables to consider when devising a control group and trying to isolate radio-loudness as the most salient parameter influencing the importance of galaxy mergers in triggering AGN.  In practice, it is not possible to control for all of these variables simultaneously.  Thus, it is still possible radio-quiet quasars which reside in unique environments are preferentially triggered by galaxy mergers.  However, it appears major galaxy mergers are an overwhelmingly essential ingredient to the triggering of AGN with powerful radio jets, and not to radio-quiet AGN at cosmic noon.

It is also important to consider the possibility of AGN jet feedback contaminating the merger classifications associated with asymmetric or disturbed host galaxy features.  There are many examples in the literature of AGN jets strongly aligned with the semi-major axis of their host galaxies, which has been dubbed the ``alignment effect'' \citep[e.g.,][]{mccarthy+87,chambers+87,dunlop+93,best+98}.  One explanation for the alignment effect is jet-host interactions as the jet propagates through the ISM and triggers star formation \citep{rees89}, thus indicating the possibility of jets altering the appearance of their host galaxies in our 3CR sample.  Additionally, it is possibile electron or dust scattering of the AGN optical/UV continuum and the nebular continuum from an extended emission line region may also contribute to the alignment effect \citep{dickson+95,tadhunter+89,deyoung89,tadhunter+02}.   We will explore the alignment effect further for the $z>1$ 3CR sample with our WFC3/UVIS~F606W (rest-frame near-UV) \textit{HST} observations in order to correlate star-formation and radio jet properties in a future study (Breiding et al., in prep.).  But, given host asymmetry is only a minor factor in our comprehensive merger criteria and it is unclear jet-host interactions have significantly affected the apparent asymmetry in our 3CR host galaxies, we do not believe jet-host interactions should substantially affect our results.  Furthermore, jet-host interactions are not expected to affect ongoing mergers or induce tidal signatures which are tell-tale signs of galaxy mergers and pervasive in our 3CR host galaxies (the \textit{HST} WFC3/IR images and PSF-subtracted images are shown in Appendix~section~\ref{sec:3CR_images} for our 3CR quasar sample).

\subsection{A Possible Jet Formation Channel: the Merger of Binary SMBHs}
\label{binaries}


One possible interpretation for our results, as first suggested by \cite{wilson_colbert_95} (also see \citealt{hughes+03}), is that major galaxy mergers can lead to radio jet formation in the so-called ``spin paradigm'' through the preceding merger of binary SMBHs and resultant spin-up of the remnant SMBH \cite[also see][]{chiaberge+11,chiaberge+15}.  In this scenario, major galaxy mergers result in the formation, and eventual coalescence, of binary SMBHs of similar mass ratio \citep[][]{begelman80}.  However, only with the right combination of initial spins and mass ratio will binary SMBH mergers spin-up the remnant black hole.  In particular, mass ratios closer to unity and highly spinning black holes aligned with the orbital angular momentum of the binary will lead to the largest spin of the remnant black hole post-coalescence.  In the case of gas-rich galaxy mergers, theoretical arguments support a scenario in which the spins of both black holes in the binary will tend to align with its orbital angular momentum through accretion torques mediated by the larger galactic-scale gas flow \citep[][]{bogdanovic+07}.  This mechanism is also supported by the observation of three nearby ($z<0.1$), young (ages $\mathrm{<~10~Kyr}$) compact symmetric objects showing jet axes normal to the galactic gas disks in near-IR/optical multiband imaging with \textit{HST} \citep{perlman+01}.  

However, ``dry mergers'' may more naturally explain the galaxy cores (i.e., missing stellar light) more frequently observed in the centers of radio-loud AGN hosts \citep[e.g.,][]{capetti+06,capetti+07}, possibly created through binary or recoiling SMBH scouring in our proposed model of jet formation \citep{begelman80,milosavljevic+01,nasim+21}.  Nevertheless, in the context of spin-powered jets such as those created through the ``Blandford-Znajek'' mechanism \citep{blandford_znajek_77}, the increase of black hole spin following a binary SMBH merger (with the right properties, as described above) would allow for the formation of the radio jet \textit{by directly tapping the increased spin energy of the remnant black hole}.  Interestingly, observations seem to show that only high mass black holes ($\mathrm{\gtrsim10^{8}M_{\odot}}$) tend to launch powerful radio jets \citep[e.g.,][]{chiaberge+11}, as supported by the 3CR black hole masses presented in this study (see Figure~\ref{fig:control_hists} and Table~\ref{table:3cr_properties}), possibly because the highest-mass black holes can achieve the highest, and most stable, spin magnitudes \citep{dotti+13}. The ``spin paradigm'' is further supported by several recent numerical general relativstic magnetohydrodynamic (GRMHD) simulations favoring spin-powered over acretion-powered black hole jets in AGN \citep[e.g.,][]{tchekhovskoy+10,tchekhovskoy+11,penna13,narayan+22}.  


Since this jet formation channel relies on a binary SMBH merger in order to spin-up the black hole and launch the jet, it makes several key predictions.  The first is that the galaxy merger responsible for triggering the radio-loud AGN should be in some stage of post-coalescence, instead of an ongoing or incipient galaxy merger.  This follows from the fact that SMBHs take some time to evolve towards the center of galaxies and eventually coaslesce themselves following a galaxy merger (binary evolution is described further in section~\ref{binary_evolution}).  While a lot of the host galaxies from our 3CR sample are classified as post-coalescence systems, there are also some number of ongoing galaxy mergers.  For these, we would predict a \textit{previous galaxy merger} is responsible for launching the radio jet in the binary SMBH merger model of jet formation we consider.  However, even though post-coalescence merger features can be observed for up to $\sim1~$Gyr \citep[e.g.,][]{lotz+08,ji+14}, this scenario is difficult to test since post-coalescence merger signatures are likely contaminated by stronger gravitational disturbances induced by the ongoing galaxy merger.  The requirement of a recent previous galaxy merger in the 3CR host galaxies classified as ongoing major mergers can naturally be explained by our radio-loud AGN residing in environments where galaxy mergers are a frequent occurance.  Indeed, observations have shown radio-loud AGN reside preferentially in galaxy overdensities (i.e., groups, clusters, and proto-clusters) \cite[e.g.,][]{venemans+07,shen+09,donoso+10,wylezalek+13,kotyla+16,ghaffari+17,retana-Montenegro+17,ghaffari+21}, which are environments conducive to frequent galaxy mergers \citep[e.g.,][]{jian+12}.    



\subsubsection{Binary SMBH Evolution \& Recoiling SMBHs}
\label{binary_evolution} 

Following the final coalescence of two galaxies in a merger, the SMBHs from each precursor system sink to the center of the merger remnant through dynamical friction \citep{chandrasekhar43}, achieving a separation of $\sim~$10~pc after a period of $\sim~100$~Myr \citep[e.g.,][]{campanelli+07,callegari+09}.  Subsequent evolution of the binary from $\sim$~10~pc to sub-pc separations is highly uncertain, relying on three-body stellar hardening \citep[e.g.,][]{sesana+07}, torqures from a circumbinary accretion disk \citep[e.g.,][]{escala+05}, and possibly three-body interactions with another SMBH \citep{hoffman+07} to remove angular momentum until the binary enters the gravitational-wave-dominated regime.  This evolutionary phase where the binary evolves from $\sim$~10~pc to sub-pc separations is referred to as the ``final pc problem'' \citep{finalpcproblem}, and can take anywhere from 10~Myr \citep[e.g.,][]{khan+15} to longer than the Hubble time \citep[e.g.,][]{yu2002} to overcome. Once entering the gravitational-wave regime (separations of $\mathrm{\lesssim ~0.01-0.1~pc}$), the binary rapidly evolves towwards its eventual coalsecence through the emission of low-frequency gravitational waves (this final stage takes $\lesssim$~tens of Myr, see e.g., Figure~1 from \citealt{begelman80}).  These gravitational waves should soon be detected by Pulsar Timing Arrays (PTAs) for SMBH binaries in the local universe \citep[][]{mclaughlin+13,hobbs+13,verbiest+16,nanograv12.5} and further out by the upcoming Laser Interferometer Space Antenna \citep[LISA,][]{lisa17} for coalescing and less massive systems.  

After binary SMBH coalescence, the remnant black hole may experience a recoil velocity-kick due to the anisotropic emission of gravitational waves \citep{peres62,bekenstein73}.  Depending on the initial spins and mass ratio of the binary, the velocity of the recoiling SMBH can be up to $\mathrm{\sim 5000~km~s^{-1}}$, although kicks of $\lesssim$ a few hundred $\mathrm{km~s^{-1}}$ are most common in simulations \citep[][]{dotti+10,lousto+12,blecha+16}.  Mass ratios near unity, highly spinning black holes, and spins somewhat misaligned with the orbital angular momentum (i.e., ``hangup kicks''), or anti-aligned and lying in the orbital plane (i.e., ``super kicks'') tend to produce the largest recoil velocities \citep[e.g.,][]{gonzalez+07,herrmann+07,lousto+11}.

Thus, another potential observable prediction of our binary SMBH merger model is the formation of radio-loud AGN in recoiling SMBH systems.  Recoiling SMBHs may manifest themselves observationally through AGN spatially offset from their host galaxy photocenters \citep[e.g.,][]{blecha+16} or the Doppler shifting of BLR emission lines \citep[e.g.,][although this can also result from a binary SMBH]{eracleous+12} as the recoiling SMBH travels through the host galaxy.  In fact, the source 3C~186 from our radio-loud AGN sample is possibly the best-evidenced recoiling SMBH in the literature, showing both a substantial 1.3~\arcsec (or 11~kpc projected) spatial offset of the quasar from its host center \citep{chiaberge+17,morishita+22} and $\mathrm{\sim2000~km~s^{-1}}$ velocity offset between its broad emission lines and host galaxy rest frame \citep{chiaberge+17,chiaberge+18,castigani+22}.  The famous, and nearby ($z=0.004$), radio galaxy M87 also shows evidence for a small $\sim$7~pc projected displacement between the AGN and host galaxy photocenter \citep{batcheldor+10}, providing some further supportive evidence for this hypothesis.  However, systematic searches (with high astrometric precision) for jetted AGN offset from their host galaxy photocenters are needed to more fully examine this hypothesis.  

Our \texttt{Galfit} models showed evidence for only one other quasar-host-center offset among the 3CR sample, a 1.4~kpc projected offset in 3C~9 (where we describe the analysis of the 3C~9 offset in detail in Appendix section~\ref{sec:3C9}).  It is possible we do not find any quasar-host-center offsets among the other $z>1$ 3CR quasars due to some combination of lower recoil velocities\footnote{As described in section~\ref{binary_evolution}, only the right combination of binary spins and mass ratio will lead to large recoil velocities post-binary coalescence.  Given that the most spin-up is experienced by remnant SMBHs where the binary spins are aligned with the orbital angular momentum, it is possible most recoil velocities in radio-loud AGN are quite small (see section \ref{binaries} where we discuss favorable binary spins for producing large recoil velocities).} and not enough time having elapsed since the recoil event to produce a detectable astrometric offset.  However, the very extreme spatial and velocity offset in 3C~186 suggests at least some cases of radio-loud AGN should lead to detectable recoiling SMBHs.  It is possible the combination of compact high-$z$ host galaxies and quasar PSF-contamination of the nuclear host galaxy light hide all but the most extreme quasar-host-center offsets\footnote{As is the case in 3C~186.  It is interesting to note that 3C~186 also has the largest size of any of our 3CR host galaxies.  Perhaps the extended nature of the 3C~186 host galaxy makes a quasar-host-center offset easier to detect.}.  One way to avoid this issue is to look for offset AGN in the type~II 3CR radio galaxies where there is no bright quasar PSF cominating the host galaxy light.  In this case, the SMBH location could be determined through high astrometric precision, very long baseline interferometry (VLBI) observations of the 3CR radio cores.  We have a completed Very Long Baseline Array (VLBA) program to do just this (VLBA/23A-297, legacy ID BB446), where the results will be reported in a future publication.

\subsubsection{The Case of the Radio-Loud Non-Merger: 3C~325}
\label{non-merger}

It is interesting to note the one radio-loud source with a clearly resolved and featureless host galaxy classified as a ``non-merger'' by our consensus classification: 3C~325.  As shown in Figure~\ref{fig:panel2}, 3C~325's host galaxy has the appearance of a smooth and featureless elliptical, as supported by its relatively high S\'ersic index of 3.4. One possible formation channel for elliptical galaxies is the merging of massive disk galaxies \citep{toomre_and_toomre_72}, as supported through simulations \citep[e.g.,][]{farouki_and_shapiro_82}.  If the binary SMBH merger model of jet formation we consider is correct, this might be a system where binary evolution took much longer than the rest of the 3CR sample and any morphological indicators of a past galaxy merger responsible for triggering the radio-loud AGN have long since dissapeared.  Better understanding this system (and others like it among radio-loud AGN samples) which appears to be an outlier among the $z>1$ 3CR sources as having no indications of any ongoing or recent major galaxy mergers will help us better understand the jet-formation physics and AGN-triggering mechanisms operating in other radio-loud AGN. 

\subsubsection{Alternative Explanations for a High Galaxy Merger Incidence among Radio-Loud AGN}
\label{selectionbias}  

While our preceding discussion focuses on the merger of binary SMBHs as the main channel for the formation of powerful radio jets through the subsequent spin-up of SMBHs in the spin-powered jet paradigm, there are other viable hypotheses not excluded by our results.  First, it is possible galaxy mergers are still important for the production of powerful jets, but it is gas accretion resulting from a galaxy merger and not the merger of a binary SMBH which acts to spin up a SMBH leading to the formation of a jet through the Blandford-Znajek mechanism \citep{moderski96a,moderski96b,moderski98, sikora+13}.  One meaningful way to observationally discriminate accretion as the mechanism for SMBH spin up as opposed to the merger of binary SMBHs is to search for evidence of recently coalesced SMBHs among radio-loud AGN.  As discussed in section~\ref{binary_evolution}, one avenue to explore in this regard is to search for evidence of gravitational-wave-recoiling SMBHs in radio-loud AGN.  

Alternatively, it is possible galaxy mergers are not important triggers for radio-loud AGN but rather the findings of high rates of galaxy mergers among radio-loud AGN samples are only a byproduct of the fact that radio-loud AGN preferentially reside in overdense group and cluster environments.  This scenario presumes that the high radio luminosities of radio-loud AGN are linked to their environments in some way other than through galaxy mergers.  One problem for this interpretation is the lack of correlations observed between cluster richness and radio luminosity in radio-loud AGN samples (see the results from the Clusters Around Radio-Loud AGN survey by \citealt{wylezalek+13} and \citealt{hatch+14}).  Another challenge for this interpretation of our results is that while radio-loud AGN are found to statistically favor over-dense environments in comparison to radio-quiet AGN, there are still a sizeable fraction of radio-loud AGN found in relatively poor environments \citep[][]{kotyla+16, croston+19}.  For instance, 3C 298 lacks any evidence for a cluster \citep[e.g.,][]{ghaffari+21} but is a well known merger with disturbed morphology (Figure \ref{fig:3cr_images2}, and \citealt{barthel+18}).  This problem can be avoided in the scenario that galaxy mergers trigger radio-loud AGN, as dense environments would only be environments conducive to galaxy mergers but not necessary by themselves to the existence of the radio-loud AGN.  One way to test the hypothesis that galaxy mergers are only a byproduct of dense environments and not important triggers for radio-loud AGN would be to control for the AGN environment in addition to the AGN and AGN host galaxy properties in future radio-loud AGN merger studies.


\subsection{The Assembly, and co-Evolution, of SMBHs and their Host Galaxies}
\label{bulge_assembly}

\begin{figure}
	\includegraphics[width=\linewidth]{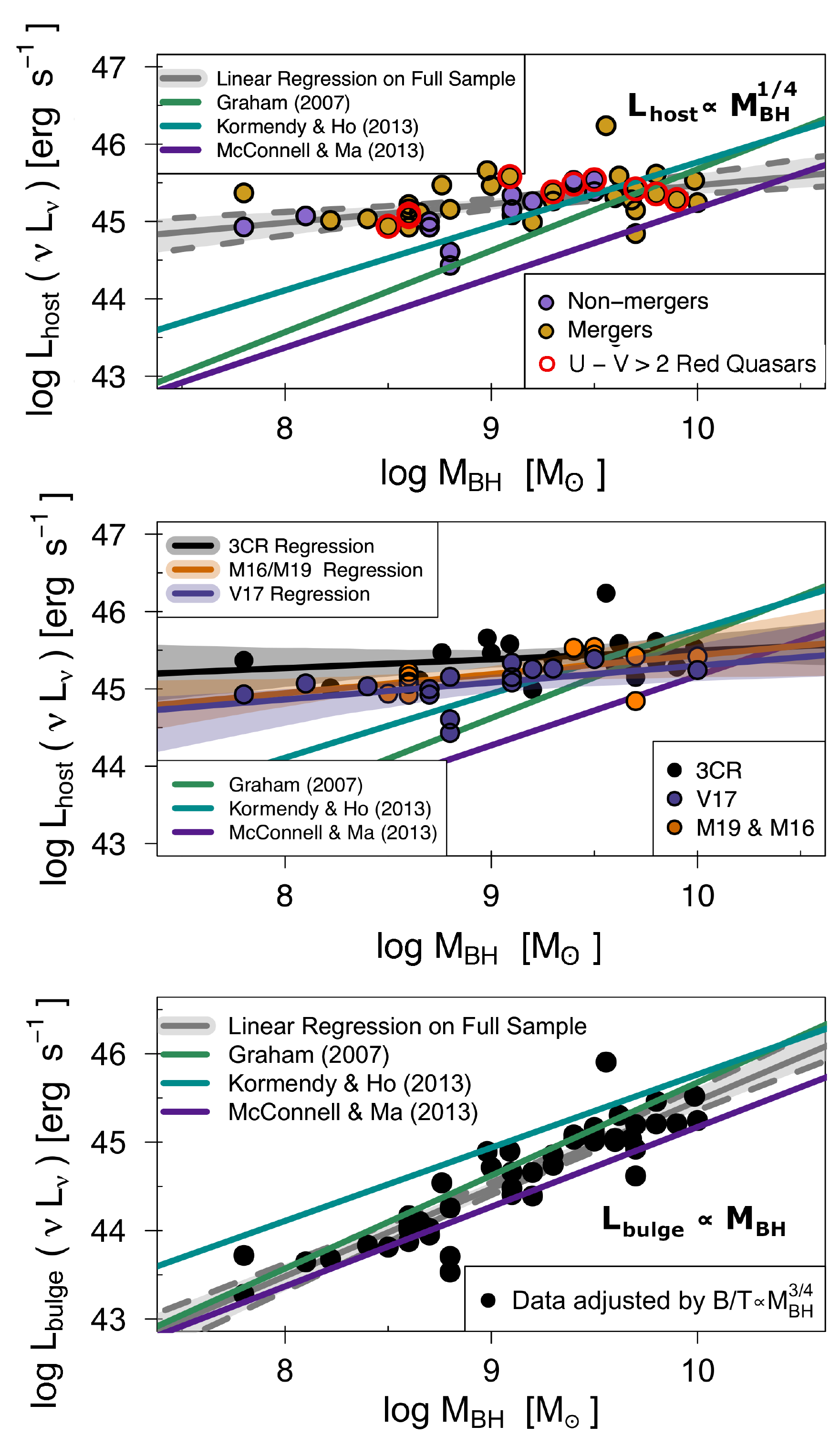}
	\caption{\textbf{\textit{Top:}} Scatter plot of R-band host galaxy luminosities and black hole masses.  Our best-fit mean linear regression line is given in dark gray, with 95\% confidence band given in gray with outlining gray dashed lines.  Expected correlation between the host spheroid component of nearby inactive galaxies with black hole mass is shown as solid colored lines (purple for \citealt{mcconnell+13}, cyan for \citealt{kormendy_and_ho_13}, and green for \citealt{graham07}).  We distinguish galaxy ``mergers'' as filled yellow circles and ``non-mergers'' as filled purple circles as determined from our consensus results.  We also mark the reddest quasars with red outlining circles, which have U-V colors in excess of 2. \textbf{\textit{Middle:}} Same as \textit{top}, but we distinguish sources based upon quasar sample.  Black circles correspond to 3CR quasars, purple V17, and orange represent a combined M19 and M16 high-$z$ sample.  Regression lines (with 95\% confidence bands) are shown for each sub-sample, with colors corresponding to the sub-sample plotted.  \textbf{\textit{Bottom:}}  Same as \textit{top} panel, but we plot host ``bulge'' luminosities after adjusting for B/T corrections following the procedure outlined in section~\ref{bulge_assembly}.  Linear regression was performed on these adjusted bulge luminosities.}
	\label{fig:mlbulge}
\end{figure}

\begin{table*}
	\centering
	\caption{Results of Linear Regression Analyses} 
	\label{table:regression}
	\begin{threeparttable}
		\begin{tabular}{lcccrcrrr} 
			\hline
			Response&Predictor&&&&&&&\\
			Variable&Variable&Sample&N$^{b}$&$\rho_{s}^{c}$&p-val$^{f}$&$\alpha$&$\beta$&$\epsilon_{0}^{d}$\\
			\hline
			$\mathrm{log\left(L_{host}\right)}$&$\mathrm{log\left(\frac{M_{BH}}{10^{9}M_{\odot}}\right)}$&full$^{a}$&48&$0.50\pm0.11$&$0.000116^{+0.0028}_{-0.000113}$&$45.22\pm0.08$&$0.25\pm0.06$&0.26\\
			$\mathrm{log\left(L_{host}\right)}$&$\mathrm{log\left(\frac{M_{BH}}{10^{9}M_{\odot}}\right)}$&3CR&21&$0.11\pm0.23$&$0.30^{+0.40}_{-0.23}$&$45.42\pm0.06$&$0.11\pm0.10$&0.26\\
			$\mathrm{log\left(L_{host}\right)}$&$\mathrm{log\left(\frac{M_{BH}}{10^{9}M_{\odot}}\right)}$&V17&15&$0.69\pm0.13$&$0.0014^{+0.011}_{-0.0012}$&$45.06 \pm0.07$&$0.24 \pm 0.12$&0.23\\
			$\mathrm{log\left(L_{host}\right)}$&$\mathrm{log\left(\frac{M_{BH}}{10^{9}M_{\odot}}\right)}$&M16/M19&12&$0.42\pm0.30$&$0.08^{+0.29}_{-0.079}$&$45.20\pm0.05$&$0.25 \pm 0.11$&0.21\\
			$\mathrm{R_{eff}}^{g}$&$\mathrm{log\left(\frac{M_{BH}}{10^{9}M_{\odot}}\right)}$&full&48&$-0.25\pm0.15$&$0.08^{+0.40}_{-0.07}$&$6.16\pm0.72$&$-2.0\pm0.97$&4.63\\
			U-V&$\mathrm{log\left(\frac{M_{BH}}{10^{9}M_{\odot}}\right)}$&full&48&$0.05\pm0.16$&$0.43^{+0.38}_{-0.32}$&$1.0\pm0.17$&$0.17\pm0.29$&1.14\\
			$\mathrm{log\left(\lambda\right)}$&$\mathrm{log\left(\frac{M_{BH}}{10^{9}M_{\odot}}\right)}$&full&48&$-0.22\pm0.15$&$0.12^{+0.45}_{-0.11}$&$-0.44\pm0.09$&$-0.40\pm0.17$&0.65\\
			Flux~Ratio$^{e}$&$\mathrm{log\left(\frac{M_{BH}}{10^{9}M_{\odot}}\right)}$&full&48&$-0.10\pm0.15$&$0.39^{+0.39}_{-0.32}$&$0.32\pm0.03$&$-0.02\pm0.05$&0.22\\
			$\mathrm{log\left(L_{bol}\right)}$&$\mathrm{log\left(\frac{M_{BH}}{10^{9}M_{\odot}}\right)}$&full&48&$0.32\pm0.13$&$0.023^{+0.16}_{-0.022}$&$46.8\pm0.1$&$0.40\pm0.18$&0.66\\
			S\'ersic~Index&$\mathrm{log\left(\frac{M_{BH}}{10^{9}M_{\odot}}\right)}$&full&21&$0.59\pm0.15$&$0.004^{+0.046}_{-0.0037}$&$2.09\pm0.18$&$1.07\pm0.29$&0.73\\
			$\mathrm{log\left(L_{bol}\right)}$&U-V&full&81&$0.22\pm0.10$&$0.047^{+0.25}_{-0.042}$&$46.8\pm0.10$&$0.08\pm0.04$&0.57\\
			\hline
		\end{tabular}
		\begin{tablenotes}
			\item[a] Full sample for $\mathrm{log\left(L_{host}\right)-log\left(M_{BH}\right)}$ correlations (and others correlations including host galaxy properties) is the combined sample of 3CR, V17, M16, and M19 quasars with host galaxy luminosities as determined from  our \texttt{Galfit} decompositions, and measured black hole masses as given in Tables~\ref{table:3cr_properties}~\&~\ref{table:control_sample_properties}.  For S\'ersic~Index$-\mathrm{log\left(M_{BH}\right)}$ correlation, the full sample includes those with measured S\'ersic indices.  
			\item[b] N is the sample size considered for the correlation.
			\item[c]  We give the mean $\rho_{s}$ and error bars defining a 68\% confidence interval as determined from our bootstrapping procedure.  The same is true for the corresponding regression coefficients.
			\item[d] The intrinsic scatter is just taken to be the standard deviation of residuals about the best-fit OLS regression line. 
			\item[e] Ratio of host to quasar flux density in the WFC3/IR \textit{HST} filter used for each quasar.
			\item[f] We report median p-values from our bootstrapped statistics, with error bars corresponding to the 68\% confidence intervals (confidence bands in Figures~\ref{fig:mlbulge}~\&~\ref{fig:mbh_vs_sersic_index} are constructed using 95\% intervals).  P-values for $\mathrm{log\left(L_{host}\right)-log\left(M_{BH}\right)}$ correlations are constructed from one-tailed hypothesis tests, and all others are two-tailed.  
			\item[g] $\mathrm{R_{eff}}$ are the effective radii in kpc.
		\end{tablenotes}
	\end{threeparttable}
\end{table*}

In Figure~\ref{fig:mlbulge}, we plot the host galaxy luminosities\footnote{We plot luminosities instead of stellar masses since we have no color information for our host galaxies and thus can not obtain reliable mass-to-light ratios.  While it is reasonable to substitute mass for host galaxy luminosity when interpreting the power-law scaling exponent, multi-band host color information and careful population synthesis modeling should be undertaken in order to obtain precise host galaxy masses.} for our quasars against their black hole masses (for all quasar host galaxies with reliable \texttt{Galfit} S\'ersic model fits as described in section~\ref{galfit}).  The host galaxy luminosities were determined at R-band since it is the closest rest-frame band for most of our quasars  (with an effective wavelength of 658~nm, and in $\mathrm{\nu L_{\nu}}$), using the magnitudes obtained from our \texttt{Galfit} decompositions and K-corrected with the elliptical host galaxy spectral template given in \cite{mannucci+01} (assuming a simple linear correction redward of the $\lambda4000$\AA~break).  We also plot the expected scalings for nearby inactive galaxy samples of ``spheroid''\footnote{Here we refer to ``spheroid'' components as either pure elliptical galaxies or the bulge component of spiral or lenticular galaxies.} luminosity components against well-measured black hole masses from \cite{graham07}, \cite{kormendy_and_ho_13}, and \cite{mcconnell+13}.  The \cite{graham07} scaling was derived from the \cite{marconi+03} K-band and  \cite{mclure_and_dunlop_2002} B-band relations (assuming R-band color corrections),  using elliptical galaxies and the bulges of disk galaxies after correcting for some discrepencies and updating source measurements from the original papers (both of the scalings are essentially the same after  the \citealt{graham07} corrections).  The \cite{kormendy_and_ho_13} and \cite{mcconnell+13} scalings are also based upon nearby samples of inactive elliptical galaxies and the bulges of disk galaxies (the latter decomposed from the total host galaxy light), color-corrected from K-band for \cite{kormendy_and_ho_13} and V-band for \cite{mcconnell+13} to R-band again using the elliptical galaxy template of \cite{mannucci+01}.    


We performed  linear regression analyses on our host galaxy luminosities and black hole masses using an ordinary least squared (OLS) estimator (the best linear unbiased estimator) for our regression coefficients, with our regression equation given by:
\vspace{-0.1mm}
\begin{equation}
	log \left(L_{host}\right)=\alpha + \beta log\left(\frac{M_{BH}}{10^{9}M_{\odot}}\right) + \epsilon_{0}~,
\end{equation}
\vspace{-0.1mm}
where $\alpha$ is the intercept, $\beta$ is the regression slope, $\epsilon_{0}$ describes the intrinsic scatter,  $\mathrm{log\left(L_{host}\right)}$ is taken to be the dependant or ``response'' variable and $\mathrm{log~\left(M_{BH}/10^{9}M_{\odot}\right)}$ is taken to be the independant or ``predictor'' variable.  We determined the strength of correlation using the non-parametric Spearman's rank-order correlation coefficient \citep{spearman1904}, $\mathrm{\rho_{s}}$
, which is less sensitive to outliers than Pearson's correlation coefficient \citep[e.g., ][]{wilcox04}.  We also performed hypothesis tests for the $\mathrm{log\left(L_{host}\right)}$-$\mathrm{log\left(M_{BH}/10^{9}M_{\odot}\right)}$ regression in order to determine significant postive correlations using a one-tailed Student's t-test \citep{zar72} (where we take p-values less than 0.05 to be cases where we can reject the null hypothesis that $\mathrm{\rho_{s}}\leq0$).  In order to determine confidence intervals for our regression, we used bootstrapping with 10,000 trials per regression analysis.  The results of our regression analyses are given in Table~\ref{table:regression}.  We find a very significant, and tight, correlation for our full sample of quasars (combined from all subsamples considered in this work), with mean $\mathrm{\rho_{s}}=0.51$ and median p-value of $0.00012$.  However, we find a much shallower slope of $\beta=0.25$ in comparison to the near-linear expectation of nearby elliptical galaxies and the bulges of disk galaxies.  As show in Figure~\ref{fig:mlbulge}, we also find no obvious trends for galaxy mergers or extremely red quasars in the $\mathrm{log\left(L_{host}\right)-log\left(M_{BH}\right)}$ plane.  In Figure~\ref{fig:mlbulge} we also show separate regression analyses after splitting our quasars into the following subsamples: 3CR, V17, and a combined sample of M16 and M19 $z\sim2$ radio-quiet sources.  As is apparent, the regression slopes are fairly consistent across the samples, although the significance of the corresponding correlations is weakened after splitting the full sample into subsamples (with all but the V17 subsample exhibiting non-significant correlations when examining the median p-values).

A near-linear scaling between black hole mass and host spheroid luminosity (and roughly spheroid mass) can be produced from models in which the host galaxy central spheroid and black hole share a common gas supply regulated by gravitational torques \citep[e.g.,][]{angles+17}, and also from ``merger avergaging'' through the hierarchical build-up of both spheroid and black hole mass through galaxy mergers \citep[e.g.,][]{peng07,jahnke+11}.  However, previous studies have also found evidence for shallower slopes in $\mathrm{L_{spheroid}-}$ $\mathrm{M_{BH}}$ scalings in the case of late-type galaxy samples \citep[ $\mathrm{M_{spheroid}\propto M_{BH}^{0.33-0.5}}$, e.g., ][]{davis+19,savorgnan+16} and wet (gas-rich) coreless galaxies possibly experiencing ``quasar''-mode non-linear growth with respect to their cold gas content \citep[$\mathrm{L_{spheroid}\propto M_{BH}^{0.30-0.45}}$,][]{graham+13}.  Although high-$z$ AGN samples seem to generally have slighly larger black-hole-to-bulge mass ratios (by a factor of $\sim2-3$, see \citealt{kormendy_and_ho_13}), their slopes are broadly consistent with local $\mathrm{L_{spheroid}-}$ $\mathrm{M_{BH}}$ relations \citep[e.g.][although see \citealt{laor01} who present evidence for the non-linear $\mathrm{M_{spheroid}\propto M_{BH}^{0.65}}$ scaling for broad-lined quasars]{bennert+11,schramm+13}.  One other possibility for our shallow slope ($\mathrm{L_{host}\propto M_{BH}^{0.25}}$)  is negative quasar feedback.  In this scenario high-$z$ quasars evolve towards the steeper local relation by the suppression of star formation as they grow (effectively moving towards the bottom right in Figure~\ref{fig:mlbulge}).  Finally, recent work has suggested that observational bias towards the most massive black holes and combined samples of ellipticals and massive bulges may have yielded near-linear correlations when separate offset, but parallel, non-linear relations (substantially shallower in $\mathrm{L_{spheroid}\propto M_{BH}}$) might apply to merger-built elliptical galaxies and the bulges of spiral galaxies \citep[][where their Figure~9 best illustrates this potential bias]{graham+23}. However, next we discuss the possibility of accounting for this discrepency in slope by our use of the total host galaxy luminosity rather than just the bulge component in the correlation.

\subsubsection{B/T Scalings and Other Considerations for $\mathrm{M_{BH}-}$Host Galaxy Correlations}
\label{bulge_scalings}

The obvious discrepancy of shallow $\mathrm{log\left(L_{host}\right)-}$ $\mathrm{log\left(M_{BH}\right)}$ slope ($\beta=0.25$) in comparison to the near-linear slopes found in samples of nearby, inactive spheroids regressed on black hole mass may also in part be explained by our use of total host galaxy luminosity instead of just the spheroid component.  Given the high-$z$ nature of our samples and PSF-contamination by our bright quasars, detailed decompositions into both disc and bulge host galaxy components is not possible for our sources.  However, if we parametrize the bulge-to-total host galaxy luminosity, $\mathrm{B/T}$, as a power-law with index of $3/4$ (i.e., $\mathrm{B/T\propto\left(M_{BH}\right)^{3/4}}$), with $\mathrm{B/T=1}$ at $\mathrm{log\left(M_{BH}\right)=10}$, then we obtain the bottom panel in Figure~\ref{fig:mlbulge} where we plot these adjusted ``bulge'' values instead of total host galaxy luminosity (we stress that these are not measured bulge values, just those inferred from the aforementioned parametrization).  This parametrization yields a consistent linear regression slope with the near-linear slopes from our expected local spheroid relations.  However, it predicts a rather extreme dependance of $\mathrm{B/T}$ on black hole mass.  In our parametrization, $\mathrm{B/T>0.5}$ (i.e., bulge-dominated) values are reached for $\mathrm{log\left(M_{BH}\right)>9.6}$ and $\mathrm{B/T=0.03}$ at $\mathrm{log\left(M_{BH}\right)=8}$.  Thus, although the most extreme black hole masses from our samples would correspond to bulge-dominated host galaxies, a majority of our host galaxies would be disk-dominated.  This result suggests a scenario where progressive major galaxy mergers lead to more bulge-dominated systems with more massive black holes, where elliptical galaxies would correspond to the most massive black holes and host galaxies with $\mathrm{B/T=1}$ \citep[e.g.,][]{hopkins+10}.  This scaling of $\mathrm{B/T}$ with black hole mass is also consistent with the positive correlations observed between S\'ersic index\footnote{Although the S\'ersic index has been shown to correlate with B/T, in general a given S\'ersic index can have a wide range of B/T values \citep[e.g.,][]{querejeta+15}.  Thus, we do not attempt to empirically constrain our $B/T$ ratios based upon our measured S\'ersic indices.} and black hole mass \citep[e.g.,][]{graham+07_sersic}.

\begin{figure}
	\includegraphics[width=\linewidth]{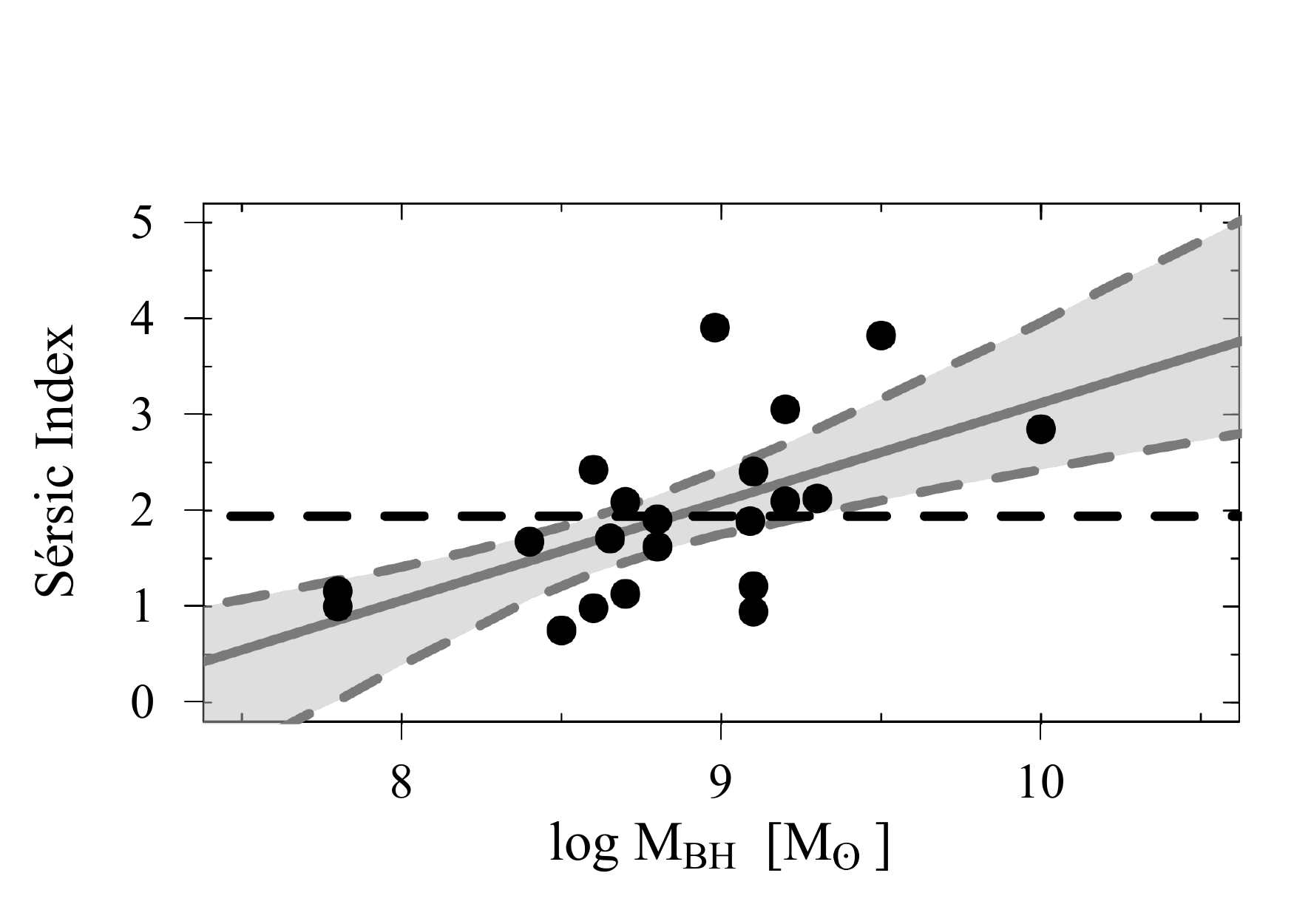}
	\caption{Here we plot the S\'ersic index against black hole mass as filled circles, with mean linear regression lines shown in dark gray and 95\% confidence bands shown with outlined dashed gray lines (constructed from the bootstrapped statistics).  A horizontal dashed black line is shown at the mean S\'ersic index of 1.94.  Regression results for this plot are given in Table~\ref{table:regression}.}
	\label{fig:mbh_vs_sersic_index}
\end{figure}

We also regressed various other source properties against black hole mass following the same methodology as outlined above (except for these we assess significance using a two-tailed hypothesis test), with the results of these analyses given in Table~\ref{table:regression}.  We found statistically significant correlations (as assessed from median p-values) only for S\'ersic index and bolometric luminosity regressed on black hole mass (where the latter correlation has also been observed previously, see e.g., \citealt{mclure_and_dunlop_2002}).  In Figure~\ref{fig:mbh_vs_sersic_index} we plot S\'ersic index against black hole mass, where there is evidently a very strong correlation consistent with the B/T scaling discussed above.  We also considered the possibility that the host galaxy luminosities are inflated at low black hole masses due to less well-resolved host galaxies and host galaxy components absorbing some of the quasar flux in our \texttt{Galfit} decompositions.  However, we would expect both smaller radii and flux ratios closer to unity for lower black hole masses if this effect were biasing our results, where no such correlations were observed.

In summary, we find the following plausible contributing factors for our shallow $\mathrm{L_{host}\propto M_{BH}^{0.25}}$ correlation: negative quasar feedback, $\mathrm{B/T}$ decreasing (substantially) towards low mass systems, gas-rich systems with highly non-linear black hole growth, or late-type galaxy morphologies which appear to follow a separate trend to strictly early-type galaxy samples.  However, we do not know which, or if any, of these factors is responsible for our observed shallow correlation.  
Better host galaxy decompositions into bulge and disc components with higher-resolution, high-sensitivity near-IR observations with the \textit{JWST} would help constrain which of these factors is most relevant in our high-$z$ quasar samples, and thus help elucidate the galaxy formation physics and evolution taking place in these systems.  Better understanding the physics and origins of different black hole mass scalings with host galaxy luminosity will aid in predictive modeling of the 
long-wavelength gravitational-wave background from nearby binary SMBHs, which is an imminent prospect in the coming years \citep[][]{nanograv12.5,chen+21}.  

\subsection{Possible Challenge to the ``Blow Out'' Paradigm}
\label{blowout}

\begin{figure}
	\includegraphics[width=\linewidth]{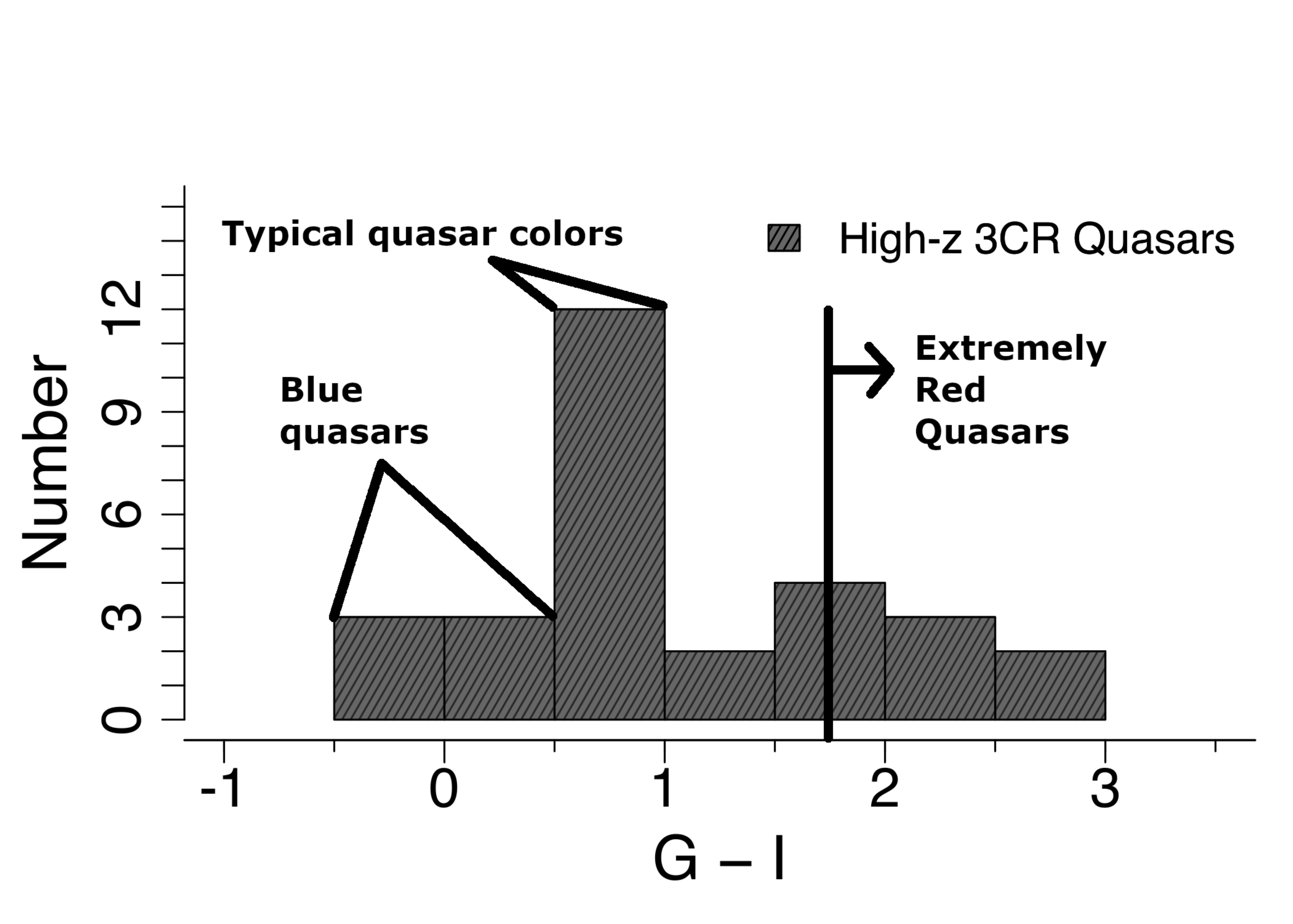}
	\caption{Histogram of rest-frame G-I quasar colors for our 3CR radio-loud AGN sample.  We also mark various regions we would expect to find blue, typical colored, or extremely red quasars as discussed in section~\ref{blowout}.}
	\label{fig:g_minus_i}
\end{figure}

There have been several recent studies of dust-reddened and obscured high-$z$ quasar samples claiming high incidences of major galaxy mergers \citep[][although see \citealt{zakamska+19} for a counter-example]{urrutia+08,glikman+15,fan+16}.  It is interesting to note the possibility of radio-loud AGN contaminating these samples and consequently inflating the observed merger fractions, where the sample from \cite{glikman+15} is radio-selected from the FIRST VLA all-sky survey (in conjunction with the 2MASS and UKIDSS infrared surveys).  Nevertheless, in the context of some semi-analytic galaxy evolution models where major (gas-rich) galaxy mergers trigger quasar activity by funneling gas to the galactic center, the quasar starts out in an early highly obscured and dust-reddened phase coinciding with early starbursts \citep[e.g.,][]{sanders88,hopkins+05,hopkins+06,hopkins+08}.  The highly obscured and reddened quasar then continues to grow in mass, and accretion luminosity, until a ``blow out'' of gas and dust by the quasar halts star formation and uncovers the (no-longer dust-reddened) blue central AGN.  This ``blow out'' paradigm in galaxy and quasar co-evolution  gained a lot of popularity as it also predicted the unambiguous association observed between ultraluminous infrared galaxies (ULIRGs) and major galaxy mergers, where in this picture ULIRGs would represent an early highly obscured quasar phase following a major galaxy merger \citep[see the review by][]{sanders+96}.  

As shown in Figure~\ref{control_samples}, our radio-loud 3CR quasars exhibit similar rest-frame U-V colors to our radio-quiet quasar control samples, where the M16 and M19 samples were constructed from the uniform color-selection algorithm presented in \cite{richards+02}.  However, in Figure~\ref{fig:g_minus_i} we also show rest-frame G\footnote{Here, we use G-band to refer to the standard SDSS \textit{g} filter.}-I quasar colors (with effective wavelengths of 477~nm and 762.5~nm for G and I-bands, respectively) determined following the same methodology as our rest-frame U-V colors (outlined in section~\ref{control_samples}).  The sharp peak in the G-I bin $[0.5,1]$ is consistent with the findings from recent color studies of local quasars with the middle 50\% of the SDSS $g^{*}-i^{*}$ color distribution in the same $0.5-1$ range \citep[][]{fawcett+21}.  Our \cite{glikman+06} near-IR optical quasar template has a G-I color of $\sim0.2$, but the authors acknowledge the small bias of a bluer continuum from this template than the SDSS~DR1 quasar composite spectrum from \cite{vandenberk+01}.  Extremely red quasar samples have been defined in previous studies using observer-frame J-K colors in excess of 1.7 \citep[i.e.,][]{glikman+04,glikman+07}, where observer-frame J-K colors roughly correspond to rest-frame G-I colors for a majority of our 3CR quasars.  Only a handful of our 3CR quasars meet this extremely red quasar color selection criteria, with many 3CR quasars consistent with normal or blue quasar colors.  However, our G-I distribution appears to have a prominent red skew, consistent with other studies finding more radio detections (many in young, compact radio jets) among red quasar samples \citep[e.g.,][]{klindt+19,fawcett+20,rosario+20,rosario+21}.

Given that a lot of our 3CR radio-loud quasars are blue 
(or representative of typical quasar colors) and are observed in recent or ongoing major galaxy mergers, our results present a potential issue for models in which blue quasars are an older evolutionary phase following the ``blow out'' phase initiated by red quasar feedback.  However, this tension can be resolved if the signatures of a major galaxy merger can last long enough for quasars to enter the blue, unobscured phase following the ``blow out'' event.   Previous studies suggest major galaxy merger signatures can last up to $\sim1~$Gyr \citep[e.g.,][]{lotz+08,ji+14}, and typical quasar lifetime estimates range from $\sim10-100$~Myr \citep[e.g.,][although estimates can range up to $\sim$1~Gyr]{martini+01,hopkins+05_qsolifetimes,hopkins+06,kelly+10}.  Thus, it would appear blue quasars may still be viable phases of quasar evolution following a ``blow out'' period if no substantial lags occur between the onset of a major galaxy merger and quasar triggering.  

\begin{figure}
	\includegraphics[width=\linewidth]{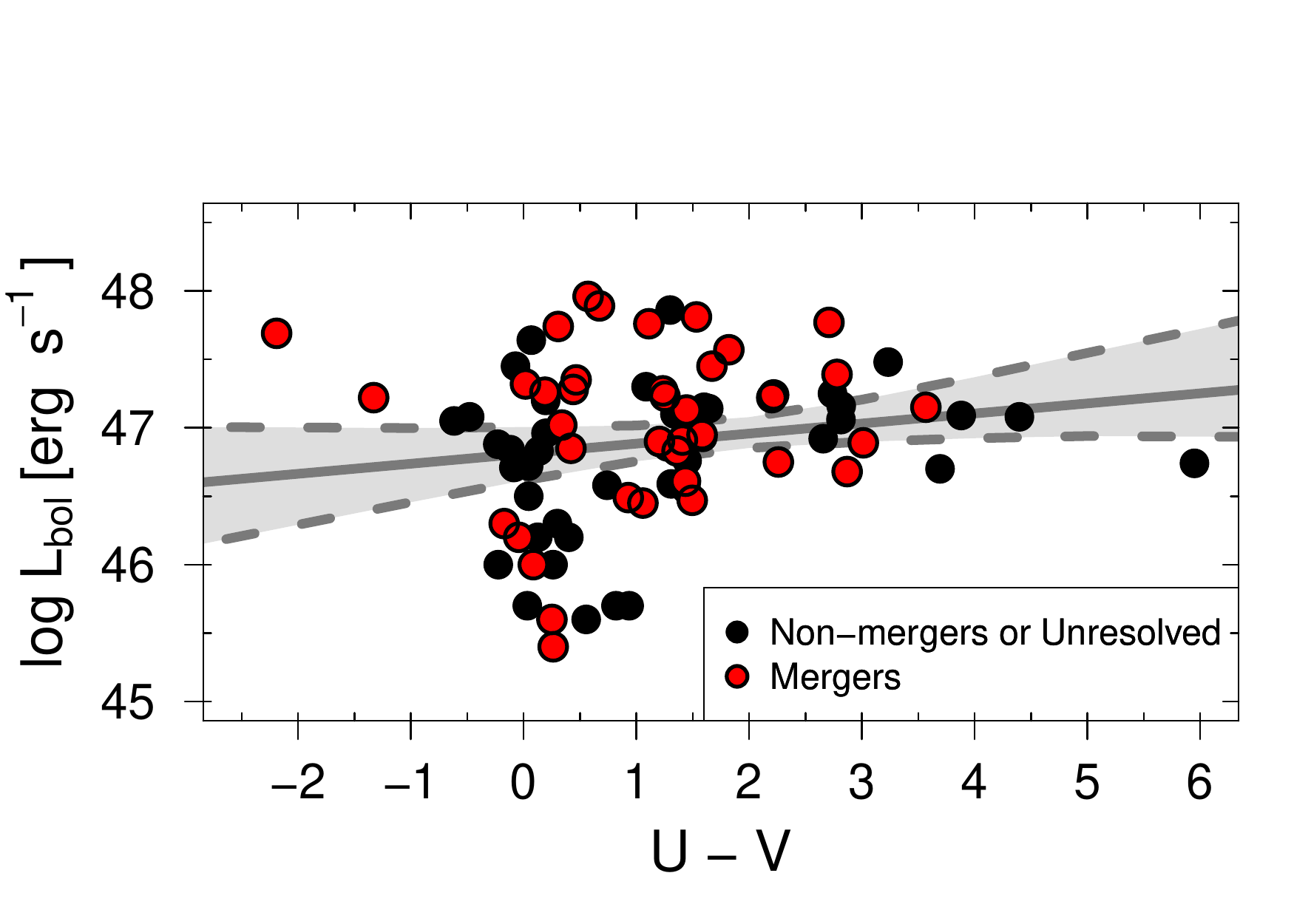}
	\caption{Plot of quasar bolometric luminosity against rest-frame U-V quasar color.  Red circles identify those systems classified as galaxy mergers by the consensus classifications (and black circles represent the ``non-mergers'').  The gray line marks the best-fit OLS regression line, with dashed lines and the shaded region outlining the 95\% confidence interval for the linear regression based upon our bootsrapped statistics.  P-values and regression coefficients are reported in Table~\ref{table:regression} for the correlation analysis. }
	\label{fig:lum_vs_color}
\end{figure}

As a further test of the ``blow out'' paradigm of galaxy evolutionary models, we plot our quasar bolometric luminosities against rest-frame U-V quasar colors in Figure~\ref{fig:lum_vs_color}, where the expectation from ``blow out'' evolutionary models is that quasar luminosities should reach a peak during the highly reddened blow-out phase \citep[e.g.,][]{hopkins+06}.  Performing a linear regression analysis following the same methodology as outlined in section~\ref{bulge_assembly} (results of which are given in Table~\ref{table:regression}), we find a signifcant positive correlation between quasar bolometric luminosity and quasar color (i.e., the reddest quasars are the most luminous), albeit with a very large degree of intrinsic scatter and shallow slope.  These findings are consistent with previous works showing red quasars tend to be more luminous than their blue counterparts \citep[e.g.,][]{glikman+12}, and also appear consistent with the ``blow out'' picture of galaxy evolution.  

\section{Summary \& Concluding Remarks}
\label{conclusion}

Using human experts, we blindly classified \textit{HST} WFC3/IR images of 28 type~I $z>1$ 3CR radio-loud quasars, in combination with three separate radio-quiet quasar control samples ($\sim~15-20$~each), for the purpose of assessing the relevance of major galaxy mergers to AGN radio-loudness.  In order to best examine the morphologies of the underlying host galaxies, we subtracted the bright quasar PSFs with \texttt{Galfit}.  Our \texttt{Galfit} decompositions included single S\'ersic models for the host galaxy components as to not oversubtract the quasar PSF, also allowing us robust host galaxy flux measurements.  

Consistent with the results for the type~II $z>1$ 3CR radio galaxies \citep{chiaberge+15}, we find nearly all of the type~I $z>1$ 3CR quasars are in recent or ongoing major galaxy mergers.  Furthermore, we find evidence for a higher merger fraction in our radio-loud quasar sample compared to all three radio-quiet quasar control samples.  
Higher-resolution, high-sensitivity near-IR observations from the \textit{JWST} may yield more robust host galaxy detections in these high-$z$ quasar samples in order to better analyze the host galaxy morphologies and obtain more reliable classifications, and thus merger fractions, for all of our samples considered (indeed the \textit{JWST} has recently detected the host galaxies of $z>6$ quasars, see \citealt{ding+22}).  However, larger sample sizes of both radio-loud and radio-quiet AGN at all redshifts would help further constrain the role of galaxy mergers towards their initial triggering.  The LOFAR Two-metre Sky Survey (LoTSS) of the northern sky at 144~MHz (with a sensitivity of $\mathrm{100~\mu Jy~beam^{-1}}$ and angular resolution of $\sim6$\arcsec) should have yielded the detection of lower-power jets at all redshifts compared to the 3CR sample of radio-loud AGN.  This would allow for more leverage to examine any dependance of merger-triggering on jet power in radio-loud AGN (although see \citealt{chiaberge+15} who find no such dependance).  To this end, the future Square Kilometer Array \citep[SKA,][]{ska} will allow more sensitive and high-resolution low-frequency radio-imaging of the southern sky.

These results support previous works finding a strong connection between radio-loud AGN and major galaxy mergers.  We hypothesize that this connection results from a scenario where binary SMBH mergers allow for relativstic jet formation through the spin-up of the remnant SMBH.  
One prediction from this model is the formation of AGN jets in recoiling SMBHs, where the magnitude of the recoil, and relative spin-up of the black hole, will depend on the configuration of binary spins prior to coalescence.  These recoiling SMBHs can be observed as spatially offset AGN (as in the case of 3C~186, and possibly 3C~9), or as AGN with velocity-offset broad emission lines.  Precise spatial offsets can be identified through follow-up VLBI observations of pc-scale radio cores tracing the SMBH positions \citep[e.g.,][]{breiding+21}.  Similarly, velocity-offset broad lines can be identified through careful modeling of their public SDSS spectra \citep[e.g.,][]{eracleous+12} or dedicated spectroscopic follow-up observations \citep[e.g.,][]{runnoe+17,chiaberge+18}.

Using our \texttt{Galfit}-measured host galaxy fluxes, we also examined the scaling between host galaxy luminosity and SMBH mass for our high-$z$ quasar samples, finding a very tight and statistically significant correlation 
with a much softer slope of $\beta=0.25$ than the near-linear slopes found in nearby spheroid relations.  This much softer relation may result from some combination of a non-linear ``quasar''-mode type of SMBH growth, the negative quasar feedback suppression of star formation, predominantly late-type galaxy morphologies, or the strong dependance of bulge-to-total (B/T) galaxy luminosity on black hole mass for our quasar samples.  Higher-resolution, high sensitivity near-IR observations with the \textit{JWST} may help resolve the bulges in our samples, thus allowing for better assessment of the host galaxy morphologies and B/T decompositions.

Finally, major merger activity in our blue radio-loud quasars presents a possible tension with galaxy evolution models in which blue quasars are the end result of red quasar feedback in the ``blow out'' paradigm.  However, this tension can be resolved if galaxy merger signatures outlive typical quasar lifetimes, as supported by recent numerical works.  We find support for the notion that red quasars are more luminous than their blue counterparts, finding a statistically significant, but high-scatter, trend between quasar bolometric luminosity and rest-frame U-V quasar color.  This correlation is consistent with the expected peak of quasar luminosity during the ``blow out'' phase when quasars are significantly dust-reddened.

\begin{acknowledgements}
We thank the anonymous referee for the many useful and constructive comments which ultimately helped improve the quality of the manuscript.  This research is based on observations made with the NASA/ESA Hubble Space Telescope obtained from the Space Telescope Science Institute, which is operated by the Association of Universities for Research in Astronomy, Inc., under NASA contract NAS 5–26555. These observations are associated with programs GO-16281, GO-13305, SNAP-12613, GO-14262, and SNAP-13023.  Some of the data presented in this paper were obtained from the Mikulski Archive for Space Telescopes (MAST) at the Space Telescope Science Institute.  The specific observations analyzed can be accessed via \dataset[DOI]{https://doi.org/10.17909/7wy0-fv26}.  This work was supported by the
HST-GO grant No. 16281.018-A.  C.O and S.A.B acknowledge support from the Natural Sciences and Engineering Research Council (NSERC) of Canada.  M.H. acknowledges DFG Program HA 3555/13-1.  
\end{acknowledgements}

\bibliography{quasar_mergers}{}
\bibliographystyle{aasjournal}

\appendix
\label{appendix}

\section{3C~119}
\label{sec:3c119_dropped}

\begin{figure}
	\includegraphics[width=\linewidth]{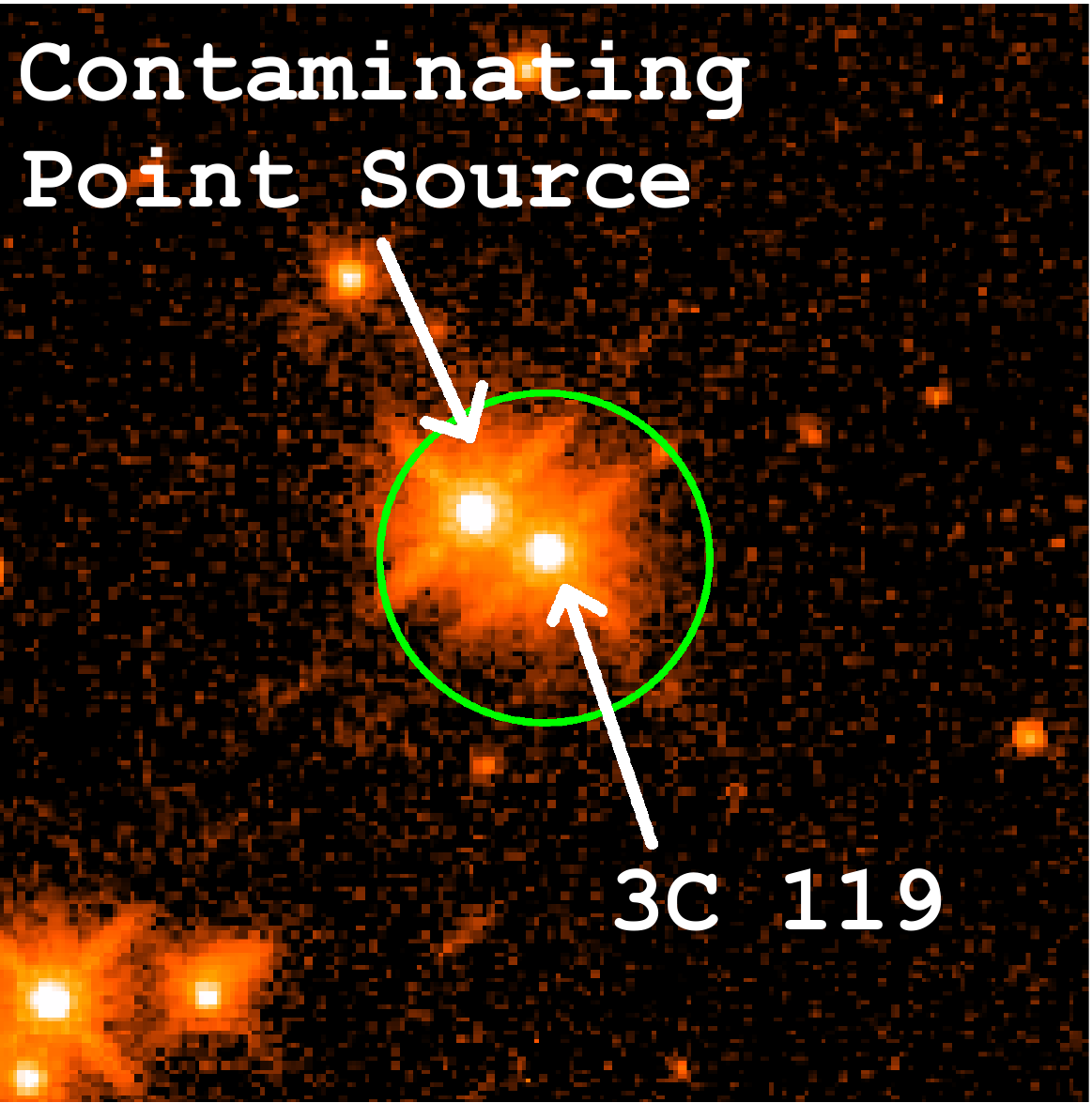}
	\caption{\textit{HST} WFC3/IR image of 3C~119, logarithmically scaled.  The green circle centered on 3C~119 has a 25~kpc projected radius at the source redshift of $z=1.02$.}
	\label{fig:3c119}
\end{figure} 

Figure~\ref{fig:3c119} shows the strong point source contamination of the quasar 3C~119.  While the contaminating point source is possibly a foreground star, it could not be matched to existing star catalogs beyond \textit{Gaia}~DR3.  In fact, both 3C~119 and the unknown point source are \textit{Gaia}~DR3 sources exhibiting proper motions on the order of 1$-$3$\mathrm{~mas~yr^{-1}}$, which may be indcative of systematic astrometric error introduced by a dual AGN \citep{hwang+20}.  Additionally, both point sources show similar IR and UV fluxes from our WFC3 IR/UVIS imaging.  Thus, we can not rule out the possibility of 3C~119 containing a double quasar without follow-up spectroscopy of its companion.  Given such a close and bright point source contamination of 3C~119, and the fact that we can not distinguish between a foreground star and a dual quasar, we chose to drop 3C~119 from our merger analysis.     

\section{PSF-subtraction uncertainties}
\label{psfsub_uncertainties}

\begin{figure*}
	\includegraphics[width=\textwidth]{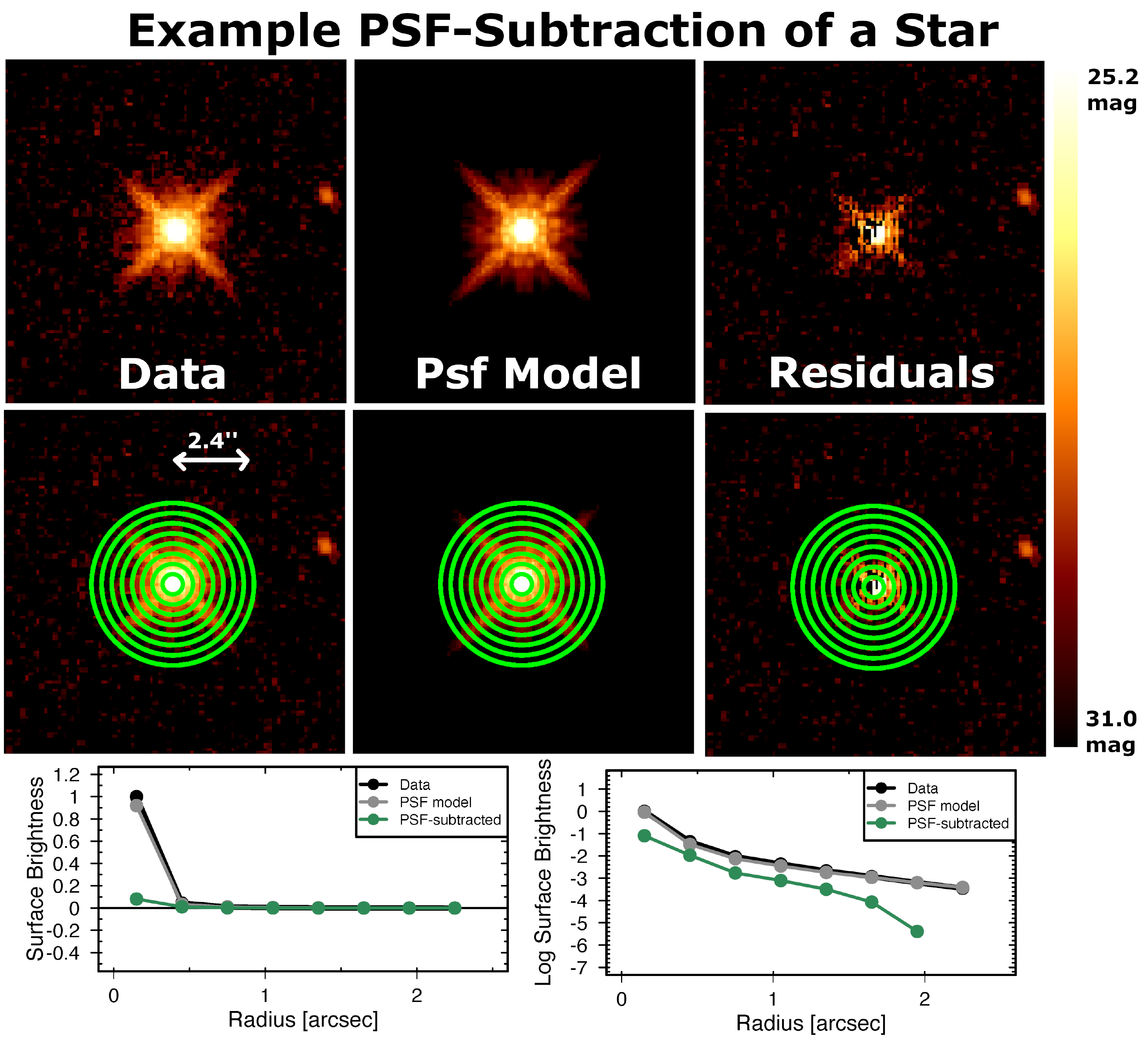}
	\caption{Here we show an example PSF subtraction of a star within the field of 3C~190.  Green circles mark concentric 0.3\arcsec spaced annuli used for the aperture photometry presented in the bottom plots.  In the bottom, we plot average surface brightness and its logarithm for each aperture defined by the concentric annuli shown in the above figures.  In black we plot the results for the image data, in gray our PSF model, and in green the PSF-subtraction results.  The surface brightness is normalized to the value found in the central 0.3\arcsec circular aperture for the original image, corresponding to $\mathrm{9.0\times10^{-27}erg~s^{-1}~cm^{-2}~Hz^{-1} arcsec^{-2}}$.} 
	\label{fig:psfsub_example}
\end{figure*}

In order to assess our PSF subtraction uncertainties, we PSF-subtracted a star in the field of the quasar 3C~190, separated from 3C~190 by $\sim$25\arcsec and with a magnitude of 18.24 (which is similar to many of our quasars in this study).  We used the same PSF-creation methodology described in section~\ref{galfit}.  Figure~\ref{fig:psfsub_example} shows the corresponding \texttt{Galfit} decomposition, as well as the average surface brightness as a function of radius in 0.3\arcsec concentric annuli as determined by aperture photometry.  There is a \textit{slight} undersubtraction of the very few center pixels, but our PSF model very closely matches all annuli past 0.3\arcsec.  
The residual flux is  $\sim8$~\% of the original flux in the center 0.3\arcsec, and $\sim20$\% past this region from $0.3-2.4$\arcsec.  Similarly, the RMS from the residual pixel flux distribution is $\sim8$\% of the PSF-subtracted total within 0.3\arcsec, and $<1\%$ from $0.3-2.4$\arcsec.  Thus, our PSF subtractions should allow for reliable recovery of extended host galaxy components beyond the most uncertain central core of the PSF.     

\section{\textit{HST} WFC3/IR Images and PSF-Subtracted Images of the $z>1$ 3CR Quasar Sample}
\label{sec:3CR_images}

\begin{figure*}
	\includegraphics[width=\textwidth]{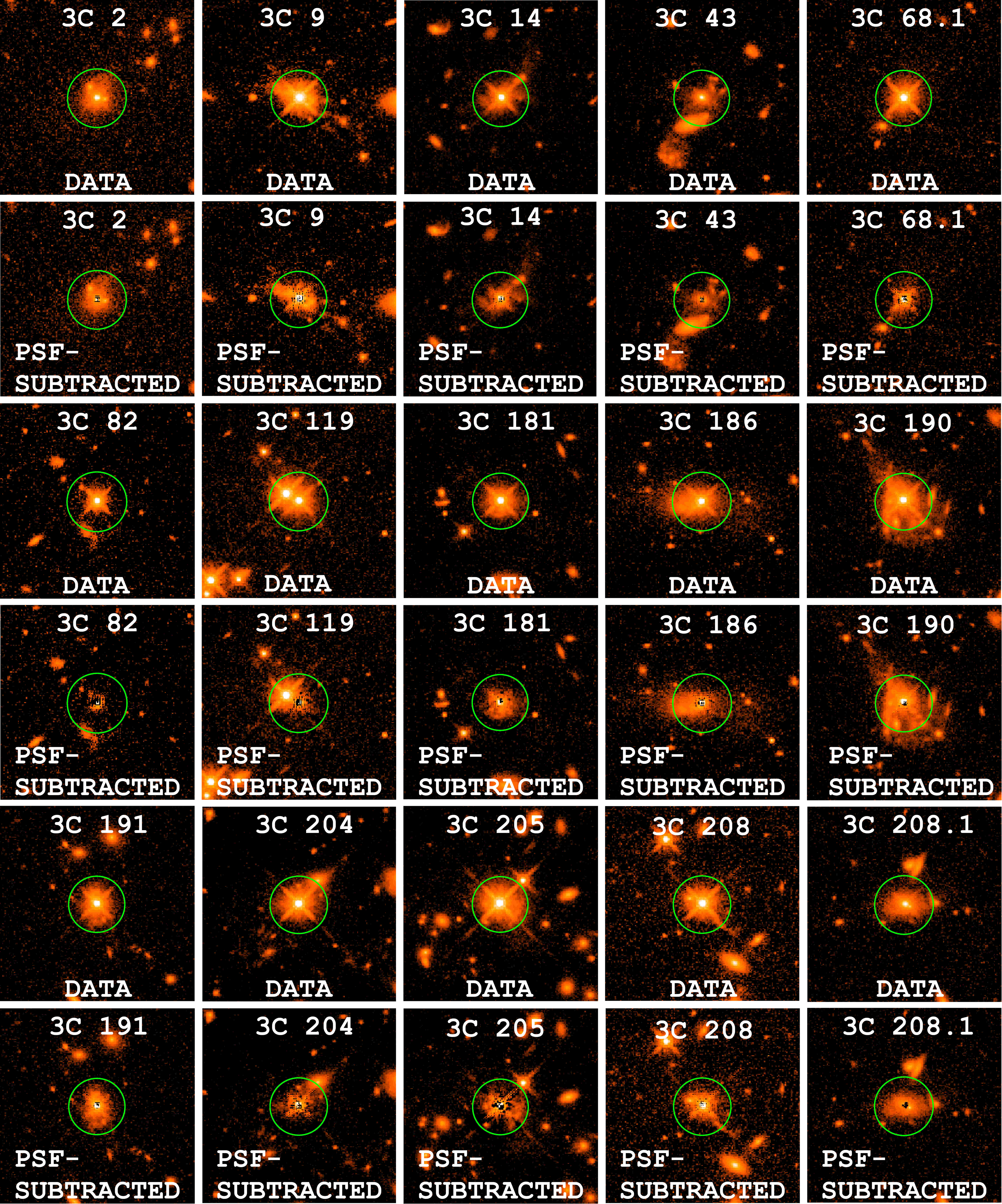}
	\caption{\textit{HST} WFC3/IR images of the $z>1$ broad-lined 3CR quasars.  These images are the same as those given in the panel decompositions used for merger classifications.  Green circles are centered on the quasars, with 25~kpc projected radii.  Images are constructed in the original detector frame, with varying right ascension/declination axis orientations.}
	\label{fig:3cr_images1}
\end{figure*} 

\begin{figure*}
	\includegraphics[width=\textwidth]{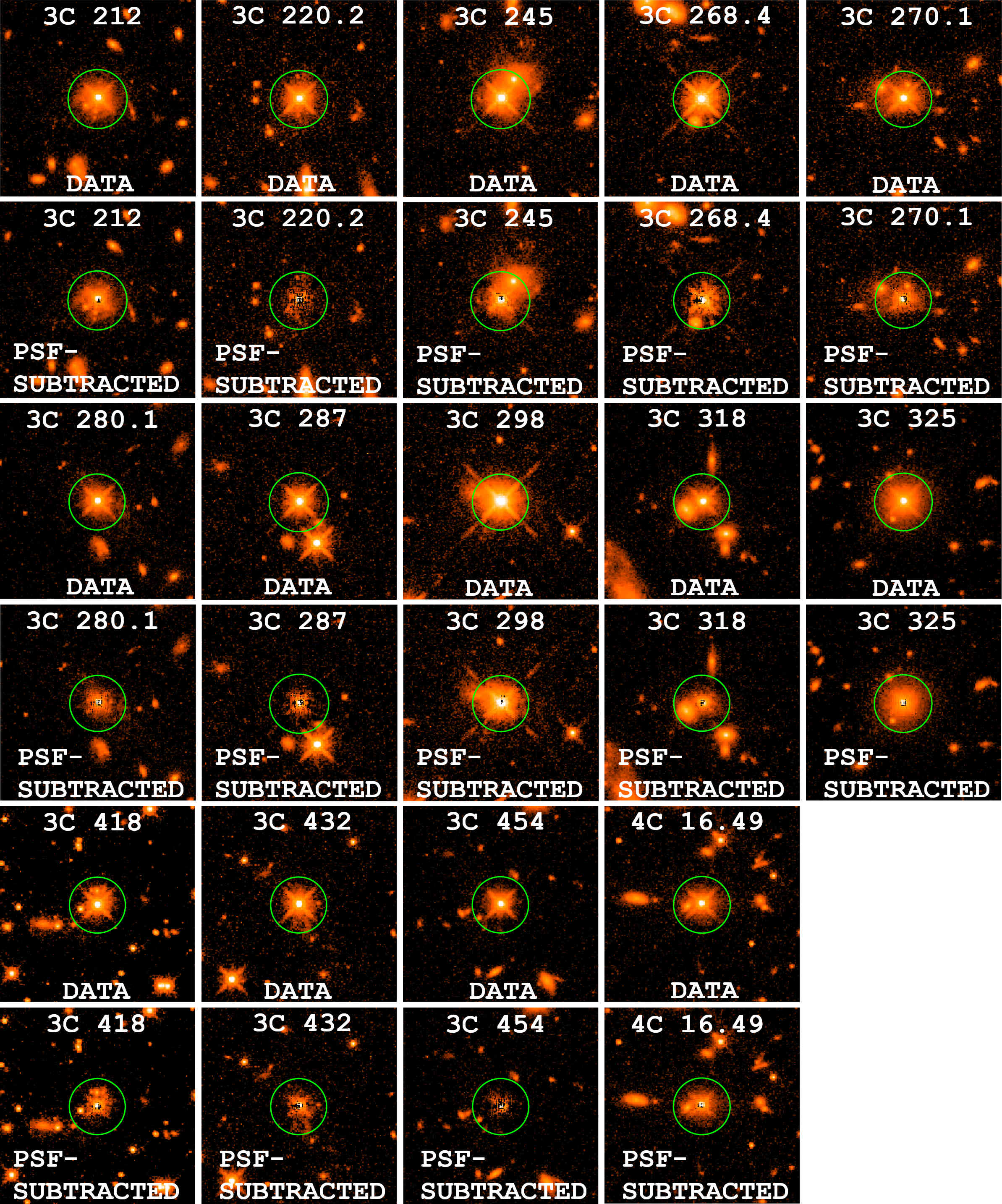}
	\caption{Continuation of Figure~\ref{fig:3cr_images1}.}
	\label{fig:3cr_images2}
\end{figure*} 

Figures~\ref{fig:3cr_images1} and \ref{fig:3cr_images2} show the \textit{HST} WFC3/IR images and PSF-subtracted images of the $z>1$ 3CR quasar sample.

\section{Modeling Tests for Blended Galaxy Components}
\label{sec:blended_components} 

\begin{figure}
	\includegraphics[width=\linewidth]{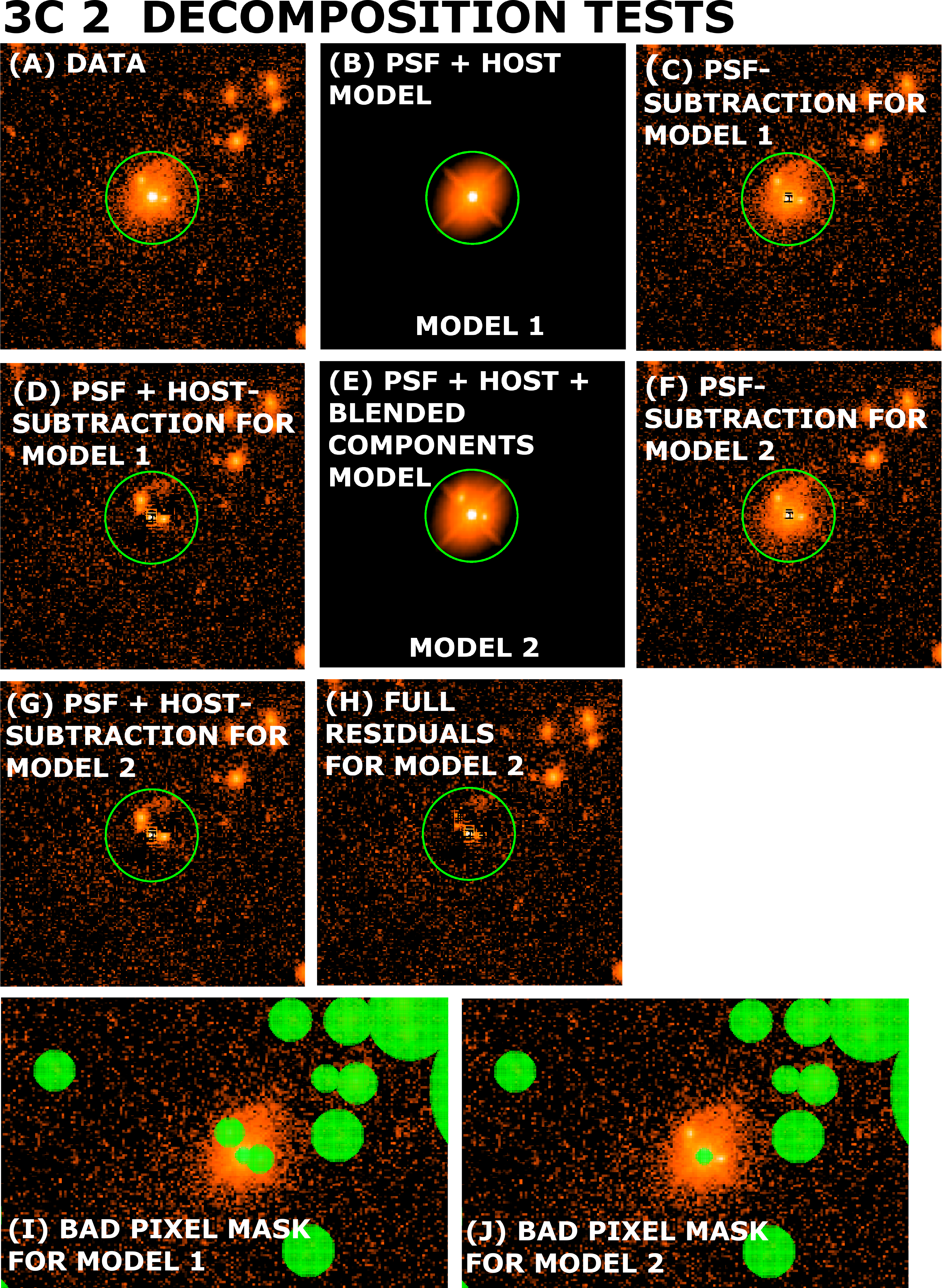}
	\caption{(A) \textit{HST}~WFC3/IR image of 3C~2.  (B) Image of the best-fit results for the model 1 components, consisting of the quasar PSF and a single Se\'rseic profile for the quasar host galaxy.  (C) PSF-subtraction for the model 1 fit.  (D) Residuals for the model~1 fit.  (E) Image of the best-fit results for the model 2 components, consisting of the quasar PSF, a single Se\'rseic profile for the quasar host galaxy, and two additional S\'erseic profiles for the blended components.  (F) PSF-subtraction for the model 2 fit.  (G) Subtraction of the quasar PSF and host galaxy S\'ersic profile model fits from the data. (H) Residuals for the model~2 fit.  (I)  \textit{HST}~WFC3/IR image of 3C~2, with bad pixels marked by green crosses for the bad pixel mask used with model 1.  (J) \textit{HST}~WFC3/IR image of 3C~2, with bad pixels marked by green crosses for the bad pixel mask used with model 2.}
	\label{fig:3c2_decomp}
\end{figure}

\begin{figure}
	\includegraphics[width=\linewidth]{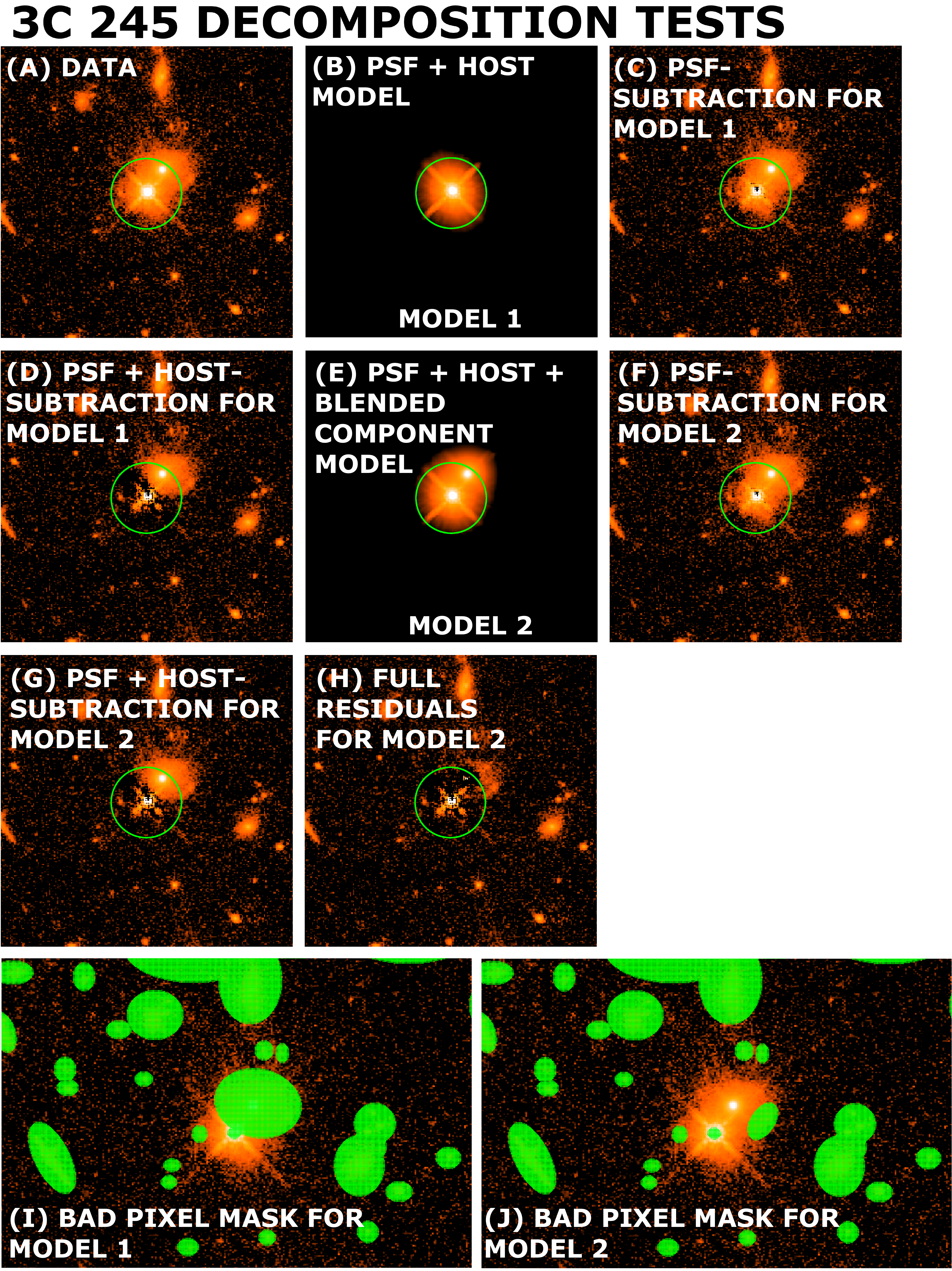}
	\caption{(A) \textit{HST}~WFC3/IR image of 3C~245.  (B) Image of the best-fit results for the model 1 components, consisting of the quasar PSF and a single Se\'rseic profile for the quasar host galaxy.  (C) PSF-subtraction for the model 1 fit.  (D) Residuals for the model~1 fit.  (E) Image of the best-fit results for the model 2 components, consisting of the quasar PSF, a single Se\'rseic profile for the quasar host galaxy, and one additional S\'erseic profile for the blended companion galaxy.  (F) PSF-subtraction for the model 2 fit.  (G) Subtraction of the quasar PSF and host galaxy S\'ersic profile model fits from the data. (H) Residuals for the model~2 fit.  (I)  \textit{HST}~WFC3/IR image of 3C~2, with bad pixels marked by green crosses for the bad pixel mask used with model 1.  (J) \textit{HST}~WFC3/IR image of 3C~2, with bad pixels marked by green crosses for the bad pixel mask used with model 2.}
	\label{fig:3c245_decomp}
\end{figure}

Here we examine the possibility of blended components other than the quasar and its host galaxy  impacting our results, given we do not include models for blended components (e.g., very nearby companion galaxies) in our \texttt{Galfit} fitting procedure.  We performed tests on two quasars from our 3CR sample, 3C~2 and 3C~245, which have host galaxies with very different types of blended light components. The 3C~2 WFC3/IR \textit{HST} data reveal two very small components within the overall envelope of the quasar host galaxy, which may be small companion galaxies enveloped by the 3C~2 host galaxy, unrelated foreground galaxies, or possibly even sites of star formation.  The 3C~245 WFC3/IR \textit{HST} data reveal a massive companion galaxy participating in an ongoing major galaxy merger with the 3C~245 quasar host galaxy (as supported by our expert votes).  

In this study, we used a single PSF and single S\'ersic profile to model the quasar and its host galaxy in \texttt{Galfit}.  We also constructed hand-made bad pixel masks for anything which did not correspond to the quasar PSF or its undisturbed host galaxy component (these bad pixel masks correspond to the pixels omitted from the \texttt{Galfit} fitting procedure).  For our tests, we performed additional \texttt{Galfit} fits for 3C~2 and 3C~245, using additional S\'ersic model components for the blended galaxy components.  For the \texttt{Galfit} fits including the blended component models we used the same bad-pixel mask as our original fits, but with the blended components unmasked.  We show the bad pixel masks we used for our original method and this blended model component method for 3C~2 and 3C~245 in Figures~\ref{fig:3c2_decomp} and \ref{fig:3c245_decomp}, respectively.  In Figures~\ref{fig:3c2_decomp} and \ref{fig:3c245_decomp}, we refer to the models using our orginal method of masking the blended components as model~1.  Similarly, we refer to the models containing additional S\'ersic profiles for the blended components as model~2.

In Figures~\ref{fig:3c2_decomp} and \ref{fig:3c245_decomp} we show \texttt{Galfit} decompositions for 3C~2 and 3C~245 using our original fitting procedure and the test fitting procedure based upon additional model components for the blended sources (models 1 and 2, respectively).  We used a fixed S\'ersic index of four for the 3C~245 host galaxy and its blended companion galaxy (in both model 1 and 2).  We also used a fixed S\'ersic index of four for the 3C~2 blended model components (in model 2).  Otherwise, we left the other fitting parameters free.  When comparing the host galaxy structural parameters recovered for 3C~2, we found a difference between the methods of 0.04 in magnitude, 0.24 in S\'ersic index, and 0.8~kpc in effective radius.  For 3C~245, we found a difference between the methods of 0.05 in magnitude and 1.6~kpc in effective radius.  The difference in magnitude between the two methods is less than the uncertainties found by  \citet[][]{simmons_urry_08} for quasar host galaxies in simulations of \textit{HST} ACS data for the GOODS survey (when comparing quasars with similar fluxes and flux ratios between the quasar and its host galaxy).  These results, along with the overall similarity in residuals between the two methods, validates the robustness of our original method of masking pixels corresponding to anything other than the undisturbed host galaxy and quasar PSF components (only including models for the quasar and its host galaxy as a single PSF and S\'ersic profile, respectively).  


\section{Quasar Host Galaxy Non-Detections}
\label{sec:unresolved_host_bias}

In Section~\ref{classifications}, we describe our classification criteria and our choice to reclassify those sources originally classified ``unresolved'' as ``non-mergers'' (corresponding to letters F and E, respectively, from the classification choices).  Given no massive galaxy companions indicative of an ongoing or incipient major galaxy merger were identified in these cases, this reclassification choice could only bias our results in the event ``post-merger'' signatures would have been present if the host galaxy were more fully resolved and detected. 

Galaxy mergers are a gravitationally violent process where stellar material can become widely distributed as the participating galaxies are coalescing, allowing for stellar material to persist beyond the \textit{HST} angular resolution limits of our study before final coalescence and dynamical relaxation.  Major galaxy mergers can also trigger galaxy-wide starbursts, significantly increasing the stellar mass and luminosity of the host galaxy as a result \citep[e.g.,][]{sanders88,elbaz+03,barnes04}.  Furthermore, gas-rich galaxy mergers in which star formation would be the most dramatic are also shown to exhibit tidal features for longer periods of time \citep{lotz+10}.  Thus, we believe both ongoing and recent major galaxy mergers are preciesly the cases where the host galaxies of our quasars, and associated merger features, are \textit{more likely} to be detected and resolved.   The above arguments not withstanding, below we assess the potential influence of host galaxy non-detections on our results.

\begin{table}
\caption{Consensus Results \& Unresolved Sources} 
\label{table:unresolved_sources}
\renewcommand\arraystretch{1} 
\centering
\begin{tabular}{ @{} l *{5}{c} @{} }
\hline
Sample&Merger&Not&Unresolved&Unresolved\\
&&Merger&&with Massive\\
&&&&Companions\\
\hline
3CR&25&3&3&0\\
V17&5&10&0&0\\
M16&5&13&9&0\\
M19&9&11&14&2\\
\hline
\end{tabular}
\end{table}

Using our human expert classifications, we define an ``unresolved'' source class to be cases where letter ``F'', corresponding to ``unresolved'' host galaxies, was voted for a majority of the time (ties broken towards F).  In Table~\ref{table:unresolved_sources}, we give our consensus results with the additional column ``Unresolved'' based upon this criteria.  The following column labeled ``Unresolved with Massive Companions'' indicates the number of objects with massive galaxy companions reclassified as ``B'', standing for incipient major galaxy mergers.  In Appendix section~\ref{sec:flux_ratios}, we describe how massive galaxy companions meeting the major-merger cutoff are determined for non-detected quasar host galaxies. 

The $z\sim2$ high-$z$ M16 and M19 quasar samples have the greatest number of unresolved sources.  In part, this can likely be attributed to a combination of cosmological surface brightness dimming \citep[][]{tolman30,tolman34} and more compact host galaxies \citep[e.g.,][]{vandokkum+08,allen+17} at high redshift.  The lower fraction of unresolved sources in the M16 sample compared to the M19 sample may be due to its high black hole mass selection criteria.  The M16 sample has much greater black hole masses than the M19 sample.  Thus, the well-known scaling between black hole mass and host galaxy stellar mass and luminosity would then predict less massive and less luminous host galaxies in the M19 sample (see \citealt{kormendy_and_ho_13} for a thorough review on this scaling relationship), as consistent with less host galaxy detections in the M19 vs M16 AGN samples.  Given the greater number of unresolved sources in both of the high-$z$ M16 and M19 control samples in comparison to the 3CR sample, below we examine the resulting potential bias.

As argued in the beginning of this section, we believe the sources classified as ``unresolved'' are most likely dominated by intrinsically non-merging galaxies.  In support of this line of reasoning, we find all six of the M19 resolved sources not fitting the ``unresolved'' class discussed above are classified as galaxy mergers.  However, we can not rule out a recent coalesced major galaxy merger that remains unresolved in these systems.  Below we consider the possibility that some fraction of our unresolved sources in the M16 and M19 control samples are actually in these post-merger systems. The mean fraction of votes for ``post-merger'' (letter D) among merger votes for each galaxy merger from our consensus is $\sim29\%$.  Assuming all of the ``unresolved'' sources from Table~\ref{table:unresolved_sources} were actually intrinsically galaxy mergers and not ``non-mergers'', we should expect $\sim$ three post-mergers in both the M16 and M19 samples.  Since the post-coalescence galaxy mergers are the only types of galaxy merger we should miss in systems with unresolved host galaxies, this assumption corresponds to increasing the number of galaxy mergers for our consensus classifications found in the M16 and M19 samples by $\sim$ three (and correspondingly three less non-mergers).  This would correspond to a merger fraction of $\mathrm{f_{m}=0.59\pm0.11}$ for the M19 sample and  $\mathrm{f_{m}=0.45\pm0.11}$ for the M16 sample\footnote{Considering the standard error, we would require a $\sim10\sigma$ deviation, or at least 80\%, in post-merger fraction among the M19 ``unresolved'' post-coalescence galaxy mergers  before the resulting merger fraction was more consistent with that obtained for the 3CR sample (again assuming all of the unresolved sources are actually in galaxy mergers).  Given the mean fraction of post-merger ``D'' votes among merger votes for the M19 galaxy mergers is 0.27, and thus consistent with that found for the rest of our samples, this scenario is highly unlikely.   However, even if all of the M16 ``unresolved'' sources were actually post-mergers, the resulting merger fraction would still be much lower than that found for the 3CR sample.}.  Both of these merger fractions are still well below that found for the 3CR sample, confirming our result of enhanced merger fraction for the radio-loud AGN in comparison to the radio-quiet AGN.  As argued in the beginning of this section, recent major galaxy mergers representative of these post-merger systems should actually be \textit{more likely} to have resolved and detected host galaxies, so we believe this potential source of bias to be insignificant and the merger fractions presented in Table~\ref{table:merger_results} are closer to the true values.  Furthermore, large-scale tidal signatures indicative of post-merger systems should still be detectable well outside of the angular resolution limits of the \textit{HST} observations presented in this study, and the most likely case is that the quasars with unresolved host galaxies from our study have not undergone a recent major galaxy merger.

\section{Criteria for Determining Host-to-Companion Flux Ratios in Unresolved Sources}
\label{sec:flux_ratios}

We used a 1:4 flux ratio throughout this work in order to distinguish between an ongoing or incipient major or minor galaxy merger.  However, this requires using our \texttt{Galfit}-modeled fluxes of the quasar host galaxies for robust estimates.  For cases in which we could not obtain reliable \texttt{Galfit} fluxes of the quasar host galaxy (as described in section~\ref{galfit}), we estimate the likely quasar host galaxy flux using the empirical scaling relation shown in Figure~\ref{fig:mlbulge} for the full set of quasars combining all subsamples.  These fluxes are then compared to any potential companion galaxies to determine if they are incipient major or minor galaxy mergers (as described in section~\ref{classifications}).

\section{The Offset Quasar 3C~9}
\label{sec:3C9}

\begin{figure}
	\includegraphics[width=\linewidth]{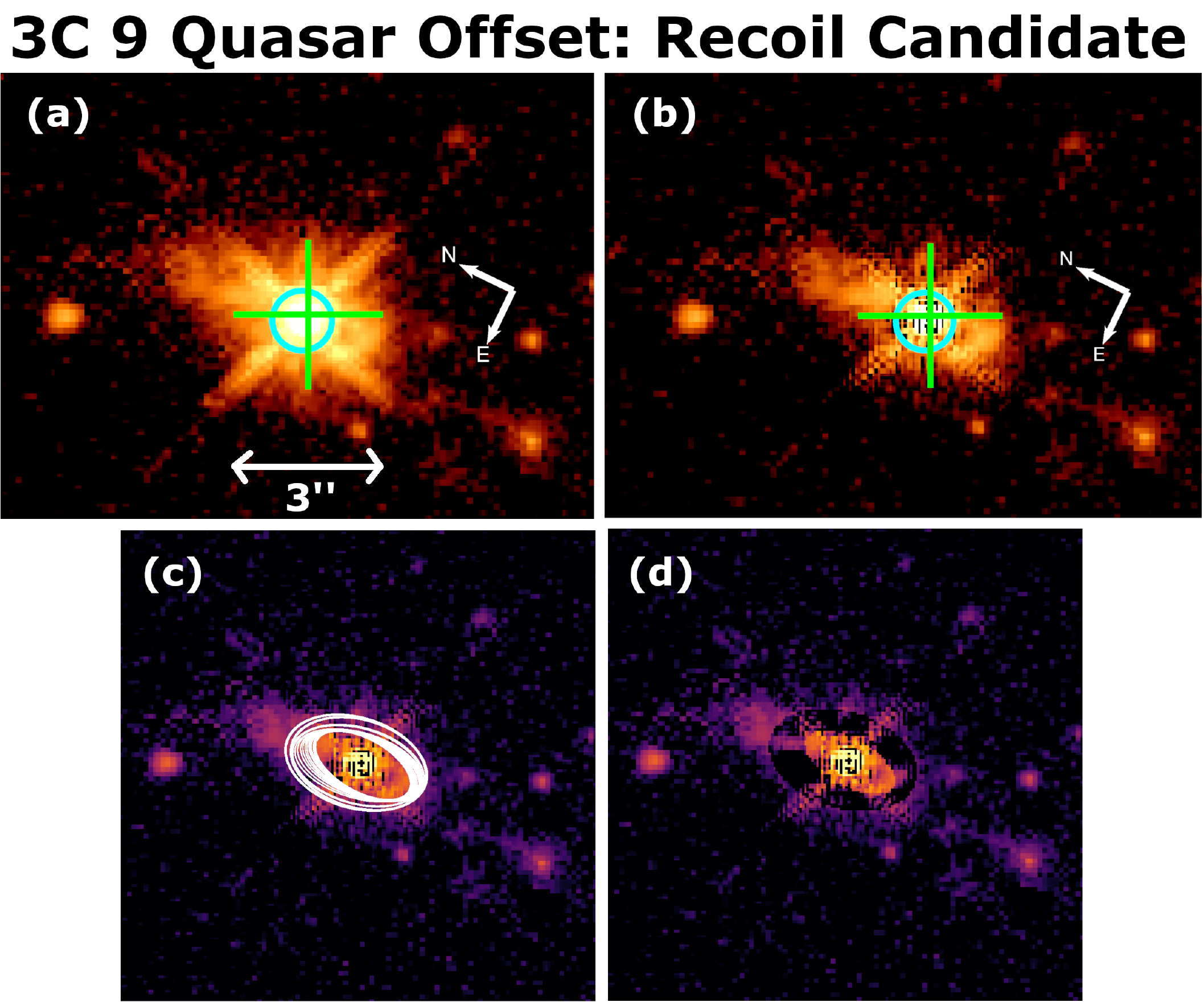}
	\caption{Image panel illustrating the quasar-host galaxy spatial offset in 3C~9, with a logarithmic flux scale for all four images. (a) \textit{HST} WFC3/IR image of 3C~9.  The compass gives the orientation of both right ascension and declination axes.  The horizontal arrow represents a 3\arcsec scale bar.  The green cross marks the host center as determined from the isophote fits shown in panel c, and the cyan circle marks the quasar best-fit position determined by \texttt{Galfit}.  (b) the same as panel a but for the PSF-subtracted image obtained from our \texttt{Galfit} decomposition.  (c) \textit{HST} WFC3/IR PSF-subtracted image of 3C~9.  Isophotes used to determine the host galaxy center are shown in white. (d) same as panel c but instead of showing isophotes used during offset analysis we show the corresponding ellipsoid model subtracted from the host galaxy light.}
	\label{fig:recoil_3c9}
\end{figure}

Figure~\ref{fig:recoil_3c9} shows the \textit{HST} WFC3/IR image of 3C~9 along with the PSF-subtracted image of its host galaxy.  Our \texttt{Galfit} decompositions yielded a best-fit  quasar PSF model offset from the the S\'ersic component by $\sim0.06$~\arcsec in the North-Eastern direction (with a 1$\sigma$ uncertainty of only $\mathrm{6~mas}$ reported by \texttt{Galfit}), or $\sim$~0.5~kpc at the redshift of 3C~9.  In order to confirm this offset, we performed isophote fits of the PSF-subtracted host galaxy image using the \texttt{photutils} \citep[][]{larry_bradley_2022_6825092} \texttt{python} package \citep[the isophote-fitting alogithrm follows the methodology of ][]{jedrzejewski87}.  We used 18 isophotes ranging from $\sim1.3-1.8$\arcsec in half-pixel increments of the semi-major axis in order to avoid the PSF-subtraction uncertainties which contaminate the center of the host galaxy.  Our best-fit isophotes are shown in Figure~\ref{fig:recoil_3c9}, where we find an $0.17~\pm~0.03$\arcsec, or $1.4~\pm~0.25$~kpc projected, quasar-host-center offset also in the North-Eastarn direction.  This roughly 5.6$\sigma$ offset is suggestive of a possible recoiling SMBH system, but it is also possible the host asymmetry resulting from a recent major merger is enough to account for this offset.  One way to test the recoil hypothesis in this source is to search for velocity-offset broad emission lines with follow-up spectroscopy.



\end{document}